\documentclass[twocolumn,trackchanges]{aastex701}
\usepackage{amsmath}
\hypersetup{linkcolor=red,citecolor=blue,filecolor=cyan,urlcolor=magenta}

\graphicspath{{./}{figures/}}
\usepackage{longtable}
\usepackage{booktabs}
\usepackage{amsmath}
\usepackage{natbib}
\begin{document}

\title{A Large Envelope Can Explain Super-Puffs}

\author{Xiumin Huang}
\affiliation{Tsung-Dao Lee Institute, Shanghai Jiao-Tong University, Shanghai, 201210, China}
\email[show]{xm\_huang@sjtu.edu.cn}  

\author{Darius Modirrousta‐Galian}
\affiliation{Tsung-Dao Lee Institute, Shanghai Jiao-Tong University, Shanghai, 201210, China}
\affiliation{School of Physics and Astronomy, Shanghai Jiao-Tong University, Shanghai, China}
\email[show]{modirrousta-galian@sjtu.edu.cn}

\author{Dong Lai}
\affiliation{Tsung-Dao Lee Institute, Shanghai Jiao-Tong University, Shanghai, 201210, China}
\affiliation{Center for Astrophysics and Planetary Science, Department of Astronomy, Cornell University, Ithaca, NY 14853, USA}
\email[show]{donglai@sjtu.edu.cn}

\author{Ren Wang}
\affiliation{Tsung-Dao Lee Institute, Shanghai Jiao-Tong University, Shanghai, 201210, China}
\email{r\_wang@sjtu.edu.cn}

\makeatletter
\gdef\@@email{%
  \parbox[t]{\dimexpr.5\textwidth-.5\columnsep-6em\relax}{%
    \raggedright\sloppy
    \nolinkurl{xm_huang@sjtu.edu.cn},\allowbreak\space
    \nolinkurl{modirrousta-galian@sjtu.edu.cn},\allowbreak\space
    \nolinkurl{donglai@sjtu.edu.cn}%
  }%
}
\makeatother

%\collaboration{all}{The Terra Mater collaboration}

\begin{abstract}
Super-puffs are a class of exoplanets with extremely low bulk densities, having masses of a few to tens of Earth masses but radii comparable to that of Jupiter. Observations suggest that these inflated radii can persist for Gyr timescales, even for planets receiving relatively low stellar irradiation. The physical mechanisms responsible for maintaining such large radii, however, remain poorly understood. Here, we model the thermal evolution of super-puffs with substantial H/He envelopes using a one-dimensional planetary structure and cooling model. We follow the coupled cooling and contraction of the planetary interior and H/He envelope while self-consistently calculating the planet’s hydrostatic and radiative--convective structure. We find that planets with H/He envelope mass fractions of approximately $40-60\%$ can retain Jupiter-like radii over Gyr timescales, provided that they begin with sufficiently hot interiors. Massive H/He envelopes substantially delay planetary contraction through two effects: (1) strong thermal blanketing, which slows the cooling of the underlying rocky core, and (2) Kelvin--Helmholtz luminosity generated by the cooling and contraction of the envelope itself, which helps maintain its high entropy and inflated radius. Our results suggest that super-puffs may represent a distinct class of planets that accreted large quantities of nebular gas during formation without undergoing runaway gas accretion.
\end{abstract}
%%\keywords{\uat{Galaxies}{573} --- \uat{Cosmology}{343} --- \uat{High Energy astrophysics}{739} --- \uat{Interstellar medium}{847} --- \uat{Stellar astronomy}{1583} --- \uat{Solar physics}{1476}}
\keywords{}

\section{Introduction} 
\begin{figure*}
{\includegraphics[width=0.99\textwidth]{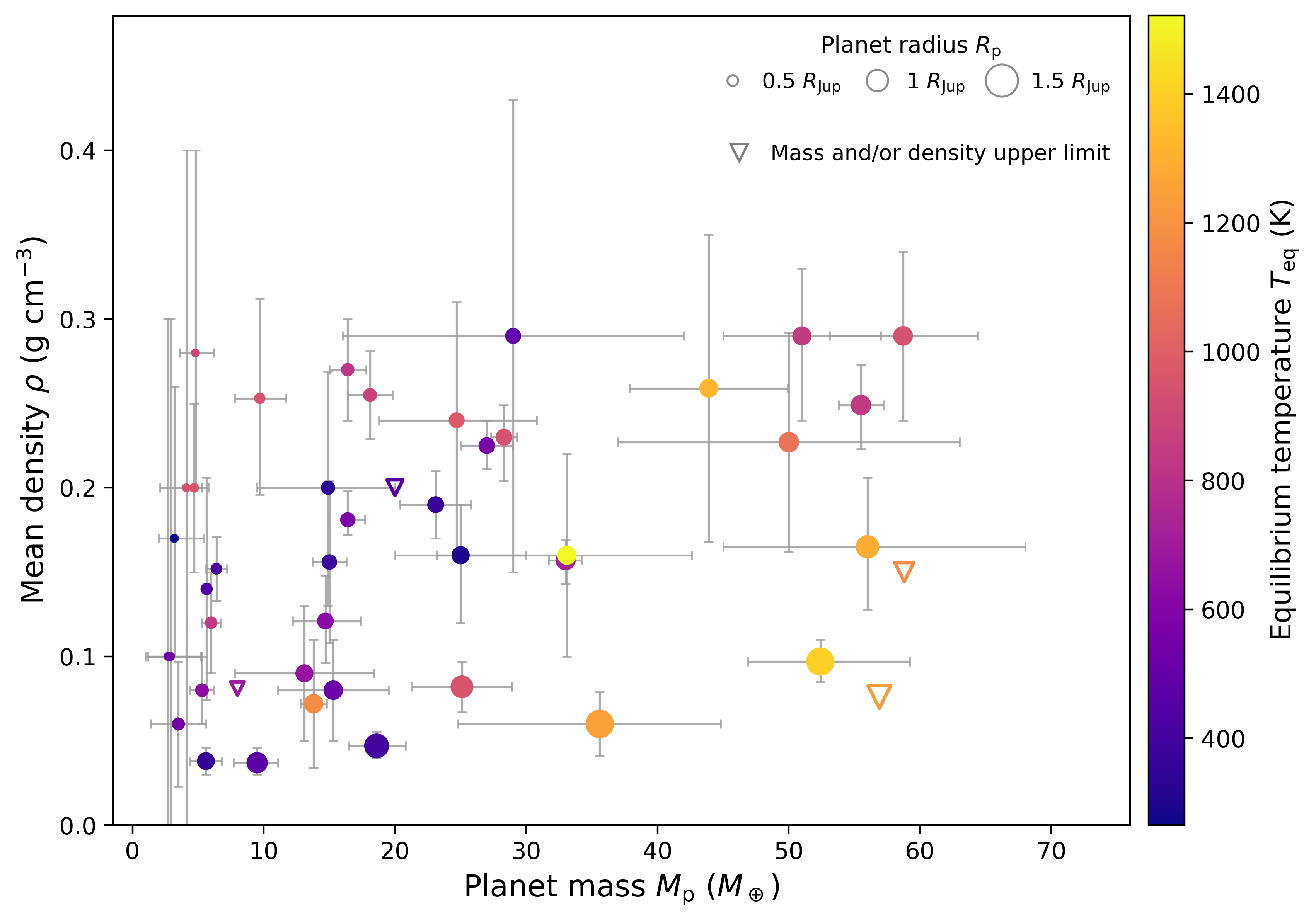}}
\caption{Masses and mean densities of super-puff candidates
with $M_p<60\,M_\oplus$ and
${\rho}<0.3\,{\rm g\,cm^{-3}}$.
The colors represent the equilibrium temperatures calculated assuming zero Bond albedo. The symbol sizes represent the planetary radii. Horizontal and vertical error bars show the uncertainties in mass and mean density, respectively. The inverted triangles indicate planets with upper limits on mass and/or density. See Appendix \ref{ap1} for the super-puff data.}
\label{fig_superpuff_population}
\end{figure*}

Observations reveal a rich diversity of planets beyond those of the Solar System. Among them, super-puffs are planets with relatively low masses but radii approaching that of Jupiter, resulting in extremely low bulk densities \citep[e.g.,][]{JontofHutter2014,Masuda2014,Barkaoui2024}. Fig.~\ref{fig_superpuff_population} shows a sample of super-puff candidates with masses below $60$ $M_\oplus$ and mean densities below $0.3$ ${\rm g/cm^{3}}$, covering a wide range of planetary masses, radii, and equilibrium temperatures \citep{Christiansen_2025}. The data of the super-puff sample are shown in Appendix \ref{ap1}. One of the most extreme examples is Kepler-51 d, which has a mass of $5.6\pm1.2$ $M_\oplus$ and a radius of $9.32\pm0.18$ $R_\oplus$, corresponding to a mean density of only $0.038\pm0.009$ ${\rm g/cm^{3}}$ \citep{Masuda2024,LibbyRoberts2026}. More recently, \citet{Dransfield2026} confirmed two Jupiter-sized planets, TOI-791 b and c, with similarly low densities of approximately $0.038$ and $0.047$ ${\rm g/cm^{3}}$, respectively, and orbital periods close to a 5:3 mean-motion commensurability. 

Two observations make the inflated radii of super-puffs difficult to explain. First, many have equilibrium temperatures of only a few hundred Kelvin, suggesting that extreme stellar irradiation is not responsible for their large radii. Second, super-puffs are also found in Gyr-old systems, such as Kepler-79 and HIP 41378 \citep{JontofHutter2014,Lund2019}, where substantial cooling and contraction should have occurred \citep{LopezFortney2014}. Indeed, the persistence of such inflated radii under these conditions exposes a fundamental gap in our understanding of planetary structure and evolution.

Proposed explanations fall into two broad categories: mechanisms that maintain large physical radii and mechanisms that increase their apparent transit radii. The former invoke massive H/He envelopes or additional internal heating to sustain large planetary radii \citep[e.g.,][]{Millholland2019,LibbyRoberts2020,Marston2026}, whereas the latter involve high-altitude hazes, dust, or circumplanetary rings that increase the observed transit radius without requiring a similarly inflated planet \citep[e.g.,][]{WangDai2019,Lu2025}. 

To maintain a large physical radius over long timescales, one possibility is a massive H/He envelope that slows planetary contraction during secular cooling \citep{Hayashi1979,MatsuiAbe1986,Calder2026}. Such envelopes provide a substantial thermal reservoir, while their outer radiative layers can limit heat loss from the interior. Interior and thermal evolution models have inferred H/He envelope mass fractions reaching tens of percent for some super-puffs \citep{LibbyRoberts2020,Marston2026}, substantially larger than the initial H/He envelope mass fractions of a few percent inferred for typical close-in low-mass planets \citep{IkomaHori2012,Wu2019}. Acquiring such massive envelopes may require favorable conditions for gas accretion \citep{LeeChiang2016,Chachan2021}, but their long-term survival may be difficult to justify if atmospheric escape driven by internal heat and stellar irradiation is efficient \citep{OwenWu2016,ModirroustaGalian2023,ModirroustaGalian2026MassLoss}. Recent studies have further shown that planetary core cooling and the extended radiative atmosphere affect the inferred envelope mass fractions \citep{Tang2025Structure}, and that substantial envelopes can survive in some super-puffs when thermal evolution is coupled to hydrodynamic escape \citep{Tang2025Escape}. Another way to sustain large physical radii is through additional internal heating such as tidal dissipation \citep{Millholland2019} and late impacts \citep{Marston2026}, which can counteract secular cooling and contraction.

In contrast, mechanisms that increase the apparent transit radius do not require the planet itself to remain as highly inflated as its observed radius would suggest. High-altitude hazes and dust carried by atmospheric outflows can increase the apparent planetary radius by making the atmosphere optically thick at low pressures \citep{WangDai2019,GaoZhang2020,OhnoTanaka2021}. Circumplanetary rings can also increase the occulting area and produce unusually large transit radii \citep{PiroVissapragada2020,Akinsanmi2020}. For example, \citet{Lu2025} proposed that migration-driven obliquity excitation could produce a tilted ring system around HIP 41378 f, allowing it to account for the planet's unusually large transit depth. However, no exoplanetary ring system has yet been conclusively detected \citep{Umetani2025}. Whereas these mechanisms can increase the apparent radius, they do not necessarily eliminate the need for a substantial H/He envelope. For example, \citet{LibbyRoberts2026} found that the JWST transmission spectrum of Kepler-51d favors high-altitude hazes, yet their thermal evolution models still require a substantial H/He envelope even after accounting for the contribution of hazes to the observed radius. 

Despite the various mechanisms that may contribute to the observed radii of super-puffs, it remains unclear whether massive H/He envelopes can remain sufficiently inflated during secular cooling without additional internal heating. In this work, we investigate this possibility using a one-dimensional planetary structure and thermal evolution model that follows the coupled cooling of a rocky interior and a substantial H/He envelope, along with the resulting contraction of the envelope. Our model self-consistently calculates the hydrostatic structure and radiative-convective temperature profile of the envelope while accounting for the thermal evolution of both the core and envelope. We adopt fixed envelope mass fractions to isolate the effects of secular cooling without accounting for atmospheric escape. Using Kepler-51 d as a representative example, we examine whether an initially hot planet can retain a highly inflated radius over hundreds of millions of years. We explore how the planetary structures depend on the equilibrium temperature, the envelope mass fraction, and the planetary mass, to investigate the conditions under which large radii can persist over Gyr timescales.

This paper is structured as follows. In Section \ref{sec_model}, we describe the planetary structure and thermal evolution model. In Section \ref{sec_results}, we present our results, including the evolution of Kepler-51 d and the dependence of planetary radii on key model parameters. We discuss our results in Section \ref{sec_discussion}, and summarize our main conclusions in Section \ref{sec_conclusion}.

\section{Model}\label{sec_model}
We consider a non-rotating planet (mass $M_p$) with a rocky core (mass $M_c$, radius $R_c$) surrounded by a hydrostatic hydrogen-helium envelope (mass $M_p-M_c$) and with no mass loss. 

The planetary structure follows mass conservation, hydrostatic equilibrium, energy transport, and local energy conservation equations,
\begin{equation}\label{eq_profile}
\begin{aligned}
\frac{dr}{dm} &= \frac{1}{4\pi r^2 \rho},\\
\frac{dP}{dm} &= -\frac{Gm}{4\pi r^4},\\
\frac{dT}{dm} &= -\frac{GmT}{4\pi r^4 P}\nabla,\\
\frac{dL}{dm} &= -T\left(\frac{\partial s}{\partial t}\right)_m.
\end{aligned}
\end{equation}
We model the chemically homogeneous H/He envelope using a fixed helium mass fraction of Y = 0.275 and the tabulated equation of state (EOS) for dense H/He mixtures \citep{Chabrier2021}. The contribution of heavy elements to the EOS is neglected. Here $r$, $P$, $\rho$, $T$, $L$, and $s$ denote the radius, pressure, density, temperature, intrinsic luminosity, and specific entropy at the enclosed mass $m$, respectively. The quantity
$\nabla \equiv d\ln T/d\ln P$ denotes the dimensionless temperature gradient. We adopt $\nabla=\nabla_{\rm rad}$ in radiative regions and $\nabla=\nabla_{\rm ad}$ in convective regions, with convective instability determined by the Schwarzschild criterion, $\nabla_{\rm rad}>\nabla_{\rm ad}$. The radiative temperature gradient is,
\begin{equation}
\nabla_{\rm rad}
=\frac{3\kappa LP}
{64\pi\sigma GmT^4},
\end{equation}
where $\sigma$ is the Stefan--Boltzmann constant and $G$ is the gravitational constant. We evaluate the opacity $\kappa$ using the analytic fit of \citet{Freedman2014}, assuming solar metallicity (\([\mathrm{M/H}]=0\)). The values of $s$ and $\nabla_{\rm ad}$ for given $P$ and $T$ are provided by the EOS of \citet{Chabrier2021}. 

The outer boundary condition is set by the global radiative energy balance between the absorbed stellar irradiation, the planet's intrinsic luminosity, and its outgoing thermal emission \citep{Guillot2010},
\begin{equation}
4\pi R_{\rm ph}^2 \sigma T_{\rm ph}^4 = L_{\rm irr} + L_{\rm int},
\end{equation}
where $L_{\rm irr} = 4\pi R_{\rm ph}^2 \sigma T_{\rm eq}^4$ is the stellar irradiation power absorbed by the planet, and
$L_{\rm int} = L(m=M_p)$ is the intrinsic luminosity evaluated at the photosphere.  Here $R_{\rm ph}$ and $T_{\rm ph}$ are the radius and
temperature at the planetary photosphere, where the photon mean free path is equal to the atmospheric pressure scale height
\begin{equation}
\frac{1}{\kappa_{\rm ph}\rho_{\rm ph}}=\frac{P_{\rm ph}R_{\rm ph}^2}{\rho_{\rm ph}GM_p}.
\label{eq:photosphere}
\end{equation}
Here $P_{\rm ph}$, $\rho_{\rm ph}$ and $\kappa_{\rm ph}$ are the photospheric pressure, density and Rosseland mean opacity, respectively, and the mass enclosed by the photosphere is regarded as the total mass of the planet $M_p$. The equilibrium temperature $T_{\rm eq}$ is defined by the balance between the absorbed stellar irradiation and the globally averaged thermal emission. For a circular orbit, it is given by,
\begin{equation}
T_{\rm eq}=T_\star\left(\frac{R_\star}{2a}\right)^{1/2}(1-A_B)^{1/4},
\end{equation}
where $T_\star$ and $R_\star$ are the effective temperature and radius of the host star, $a$ is the orbital semi-major axis, and $A_B$ is the Bond albedo. The Bond albedo depends on atmospheric properties, such as composition and cloud formation, and can vary substantially among planets \citep{Sudarsky2000,Cahoy2010,CowanAgol2011}. Nevertheless, for an illustrative range of $A_B=0$ -- $0.6$, the equilibrium temperature varies by at most approximately $20\%$ owing to its weak dependence on $(1-A_B)^{1/4}$. We therefore expect uncertainties in $A_B$ to have limited effect on our conclusions regarding the long-term radius evolution of super-puffs.

The rocky core likely consists of a metallic iron inner core plus a convective silicate magma ocean (also called molten silicate mantle). For the radius of the rocky core, which serves as the lower boundary of the envelope, we adopt the following standard mass--radius relation \citep{Seager2007,Zeng2016,LopezFortney2014},
\begin{equation}
    R_c=R_\oplus\left(\frac{M_c}{M_\oplus}\right)^{1/4},
\end{equation}
where $R_c$ and $M_c$ are the rocky core radius and mass, respectively, and $R_\oplus$ and $M_\oplus$ are the Earth's radius and mass. At the core-envelope boundary, we impose $m=M_c=(1-f_{\rm env})M_p$ and $T=T_c$, where $f_{\rm env}$ is the envelope mass fraction and $T_c$ is the magma-ocean temperature.

We model magma-ocean cooling by balancing its rate of thermal energy loss with the luminosity supplied to the envelope. We neglect heat transfer from the metallic core to the magma ocean because the compositional discontinuity at the metallic iron-silicate boundary prevents direct convective exchange, restricting heat transfer primarily to conduction, which is inefficient \citep{DahlStevenson2010,zhou2024,ModirroustaGalian2026}. Magma cooling is therefore modeled through,
\begin{equation}\label{eq_cooling}
M_{\rm c}(1-f_{\rm iron})c_p\frac{dT_{\rm c}}{dt}=-L_{\rm c},
\end{equation}
where $f_{\rm iron}$ is the metallic iron mass fraction of $M_c$ and $L_{\rm c}$ is the core luminosity, supplied by the magma-ocean cooling. The specific heat capacity of the magma ocean, $c_p$, is evaluated using the temperature- and pressure-dependent relation \citep{Korenaga2023,ModirroustaGalian2025},
\begin{equation}\label{eq_cp}
c_p=6.27\times10^6+4.11\times10^3 T_{c}-2.11\times10^{-7}P,
\end{equation}
where all quantities are in cgs units, and $P$ is the pressure at the base of the gaseous envelope. Throughout this work, we adopt $f_{\rm iron}$ = 0.3; this amounts to reducing the effective heat content of the whole rocky core by $30\%$ --- we note that the heat capacity of the magma ocean (Eq. \ref{eq_cp}) is likely uncertain by a factor of 2. The uncertainty in $f_{\rm iron}$ does not affect our results in any appreciable way.\footnote{As discussed in Section \ref{sec_Lc},  even when the core luminosity is set to zero, the thermal evolution of the planet is hardly affected if the envelope is sufficiently large.}

We solve Eqs.~(\ref{eq_profile}) and (\ref{eq_cooling}) subject to the inner and outer boundary conditions to determine the planetary structure and thermal evolution for specific $M_p$, $f_{\rm env}$, $T_{\rm eq}$, and initial $T_c$. We follow the thermal evolution by specifying a sequence of decreasing core temperatures rather than fixed time steps. At each step, we integrate the envelope structure and calculate the entropy change relative to the previous profile to determine the envelope's contribution to the luminosity. The time interval is determined from Eq.~(\ref{eq_cooling}), and the envelope structure and cooling time are solved iteratively to satisfy the envelope mass and luminosity boundary conditions. Calculating the atmospheric entropy change requires the solution from the previous step, which is unavailable at initialization. We therefore discard the first five steps to allow the numerical solution to converge.

In principle, a thermal boundary layer could exist between the magma ocean and the gas envelope \citep[e.g.,][]{ModirroustaGalian2023}. However, we found that for the large envelope fraction ( $\gtrsim$ a few percent)
of interest in this paper, the temperature jump across the boundary layer is small ($\Delta T/T_c\ll 1$) and the effect of this boundary layer is negligible.

\section{Results}\label{sec_results}
In this section, we take Kepler-51 d, one of the lowest-density known super-puffs, as a representative case to investigate the conditions required to reproduce its observed radius of 9.3 $R_{\oplus}$ at its current age 500 Myr \citep{Masuda2024}. We explore the effects of the equilibrium temperature, the H/He envelope mass fraction, and the total planetary mass. We then extend our analysis to a wider range of these parameters.

\subsection{Sample Snapshots}
\begin{figure*}
{\includegraphics[width=0.99\textwidth]{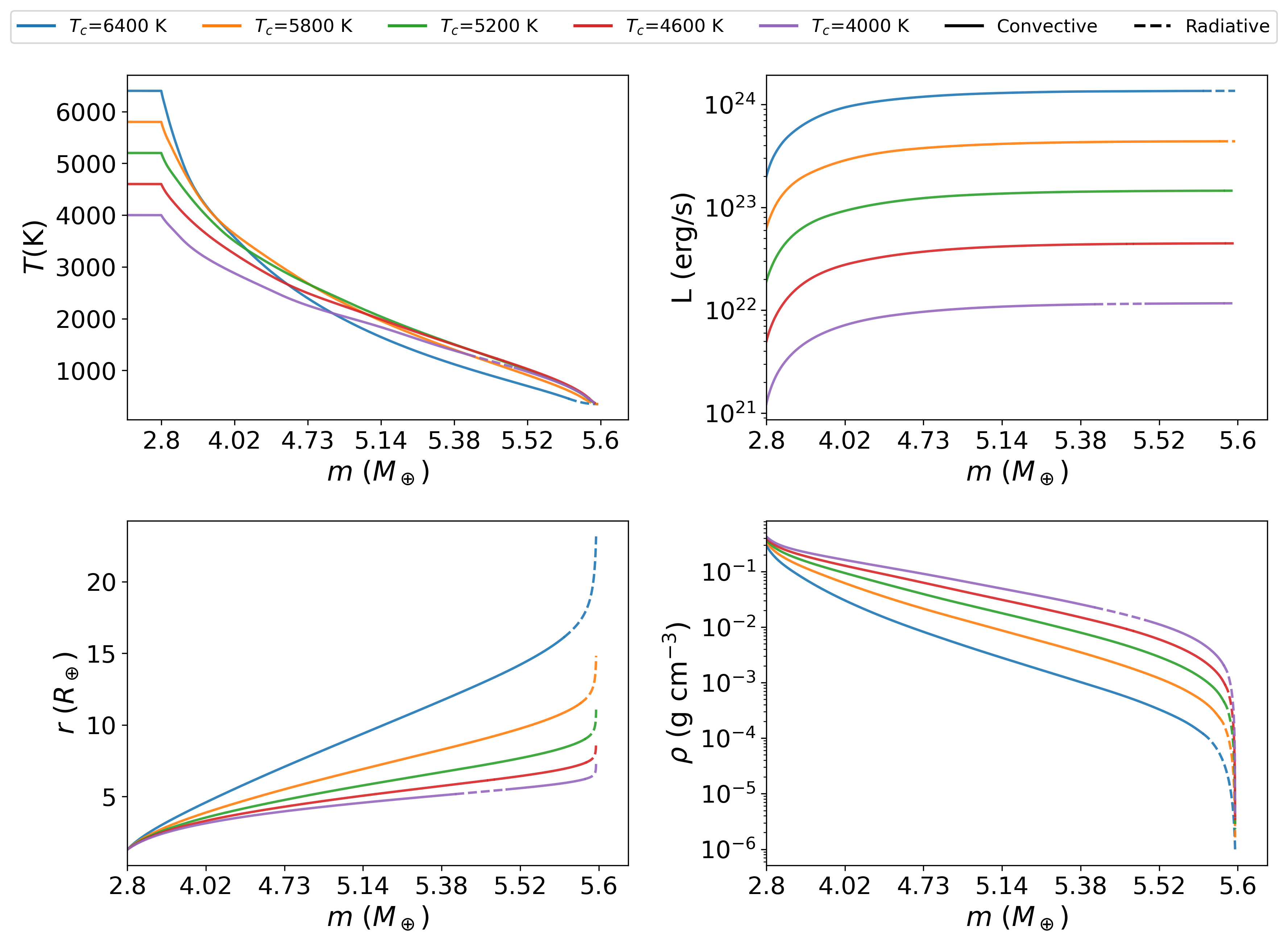}}
\caption{Planetary structure profiles for Kepler-51 d with
$M_p=5.6\,M_\oplus$, $T_{\rm eq}=350$ K
(assuming zero Bond albedo), and an assumed
H/He envelope mass fraction of $f_{\rm env}=0.5$.
The upper-left, upper-right, bottom-left, and bottom-right
panels show the temperature, luminosity, radius,
and density profiles as functions of the enclosed mass $m$,
respectively. Different colors correspond to different core
temperatures $T_c$. The solid and dashed lines represent regions where energy is transported by convection and radiation, respectively. The core-envelope boundary is located at
$m=M_c=2.8\,M_\oplus$.}
\label{fig_profile}
\end{figure*}

Adopting the observed mass of Kepler-51 d, $M_p=5.6\,M_{\oplus}$, and an equilibrium temperature of $T_{\rm eq}=350$ K (assuming zero Bond albedo), we calculate the planetary structure assuming an H/He envelope mass fraction of $f_{\rm env}=0.5$. Fig.~\ref{fig_profile} presents the temperature, intrinsic luminosity, radius, and density profiles for different core temperatures $T_c$. As shown by the solid and dashed lines in Fig.~\ref{fig_profile}, in general the inner envelope is convective, whereas the outer envelope is radiative. As $T_c$ decreases, the mass of the outer radiative envelope decreases, and an inner radiative zone may also develop in the middle of the convection zone (see the $T_c=4000$ K curve).

Increasing $T_c$ causes the envelope to expand, allowing the photospheric radius to match or exceed the observed 9.3 $R_\oplus$ of Kepler-51 d while remaining well below the Hill and Bondi radii, both of which exceed 200 $R_\oplus$ for the adopted parameters (see Eqs. (10) and (32) in \citet{DAngelo2010} for the definitions of these two radii). The upper-right panel of Fig.~\ref{fig_profile} also shows that the cooling and contraction of the envelope contribute substantially to the intrinsic luminosity, alongside heat released by the magma ocean.

\subsection{Thermal Evolution}\label{sec_evo}
The thermal evolution of super-puffs depends primarily on $T_{\rm eq}$, $f_{\rm env}$, and $M_p$. In the following, we investigate how these parameters affect the long-term cooling and contraction of the planet.

\subsubsection{Dependence on the equilibrium temperature}
Because the albedo of Kepler-51 d is poorly constrained, we consider equilibrium temperatures of $T_{\rm eq}$ = 350 K and 278 K, corresponding to Bond albedos of 0 and 0.6, respectively. For comparison, we also explore two lower equilibrium temperatures, $T_{\rm eq}=200$ K and $111$ K, to investigate the thermal evolution under weaker stellar irradiation. We fix the total planetary mass to $M_p=5.6\,M_{\oplus}$ and the envelope mass fraction to $f_{\rm env}=0.5$.

At sufficiently high $T_c$, the initial cooling is rapid, making the subsequent long-term evolution relatively insensitive to the exact initial $T_c$. We therefore adopt a high initial $T_c$ and verify that further increasing its value has a negligible effect on the predicted planetary radii at the ages of interest.

\begin{figure*}
{\includegraphics[width=0.99\textwidth]{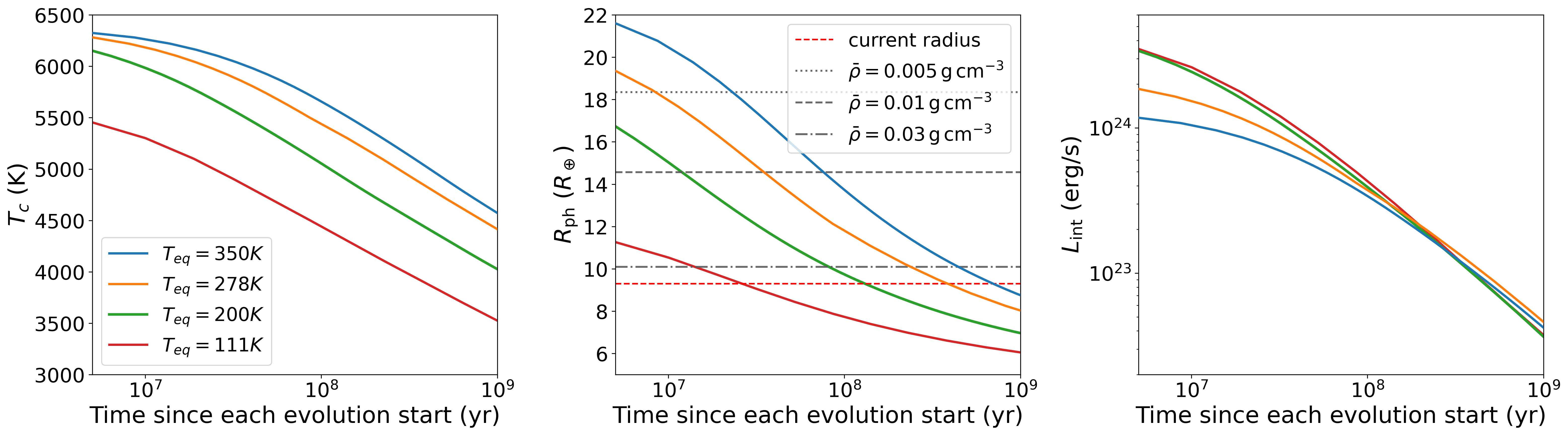}}
\caption{Thermal evolution of a planet with $M_p=5.6\,M_\oplus$ and an H/He envelope mass fraction $f_{\rm env}=0.5$ for different equilibrium temperatures
$T_{\rm eq}$. The left, middle, and right panels show the evolution of the core temperature $T_c$, photospheric radius
$R_{\rm ph}$, and intrinsic luminosity at the photosphere $L_{\rm int}=L(m=M_p)$,
respectively. Different colors correspond to different $T_{\rm eq}$, as indicated in the legend. For Kepler-51d, $T_{\rm eq}=350$ and $278$ K correspond to Bond albedos of 0 and 0.6, respectively, and the lower temperatures 200 K and 111 K are considered for comparison. The horizontal red dashed line marks the observed radius of Kepler-51d ($9.3\,R_\oplus$).
The gray dotted, dashed, and dash-dotted lines represent
radii corresponding to mean densities of $0.005$, $0.01$, and $0.03\,{\rm g\,cm^{-3}}$, respectively, for the adopted planetary mass.}
\label{fig_Teq}
\end{figure*}

Fig.~\ref{fig_Teq} shows the evolution of the core temperature $T_c$,
photospheric radius $R_{\rm ph}$, and intrinsic photospheric luminosity $L_{\rm int}$ for different $T_{\rm eq}$. The horizontal red dashed line in the middle panel marks the currently observed radius of Kepler-51 d, $R_p=9.3\,R_{\oplus}$. As the planet evolves, both $T_c$ and $R_{\rm ph}$ decrease together. Whereas the planetary radius at a given $T_c$ depends only weakly on $T_{\rm eq}$, the cooling and contraction timescales are strongly affected by stellar irradiation. As shown in the right panel, a higher $T_{\rm eq}$ corresponds to a lower intrinsic photospheric luminosity during the early evolution, indicating a slower release of internal energy.
Stronger stellar irradiation provides a more effective thermal
blanket, suppressing the loss of internal energy, thus slowing planetary contraction. Consequently, planets with higher equilibrium temperatures can maintain their inflated radii over longer timescales. Given the estimated age of Kepler-51 d of approximately 500 Myr \citep{LibbyRoberts2020}, our model, with $T_{\rm eq}=278$ K to 350 K, predicts a planetary radius comparable to the observed $9.3 \,R_{\oplus}$, demonstrating that a massive H/He envelope can sustain the highly inflated structure of Kepler-51 d over its current lifetime.

\subsubsection{Dependence on the envelope mass fraction}
We consider $f_{\rm env}=0.3$, $0.4$, $0.5$, and $0.6$, while fixing the total planetary mass to $M_p=5.6\,M_{\oplus}$ and the equilibrium temperature to $T_{\rm eq}=350$ K.

\begin{figure*}
{\includegraphics[width=0.99\textwidth]{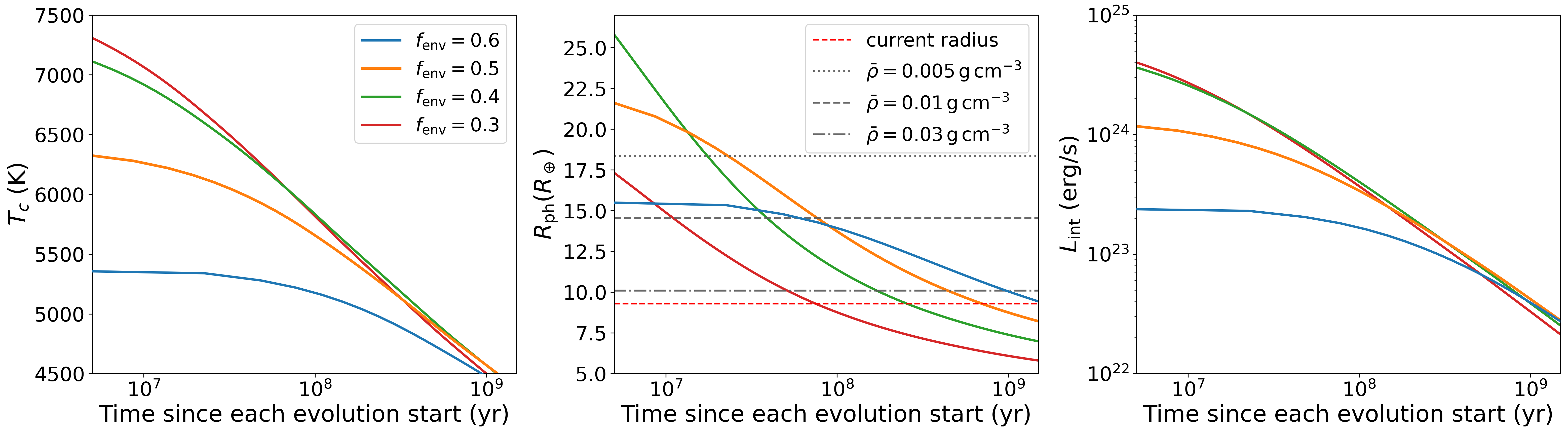}}
\caption{Same as Fig. \ref{fig_Teq}, but for different envelope mass fraction, with $M_p=5.6M_{\oplus}$ and $T_{\rm eq}=350$ K.}
\label{fig_fenv}
\end{figure*}

Fig.~\ref{fig_fenv} shows the thermal evolution for different envelope mass fractions. As shown in the left panel, the core temperatures gradually converge, with similar cooling rates, $dT_c/dt$, at later stages of evolution ($t\gtrsim 5\times10^8$ yr). In contrast, the planetary radii are significantly different, with larger envelope mass fractions producing more extended structures because of the weaker gravitational attraction exerted by their smaller cores. Thus, for the same planetary mass, the strong dependence of planetary radius on $f_{\rm env}$ is primarily associated with the internal structure rather than the core cooling rates. The envelope itself also contributes to the planet's intrinsic luminosity through its own cooling and contraction. Consequently, planets with larger envelope mass fractions remain more inflated and can sustain their large radii over longer timescales.

At the estimated age of Kepler-51 d, models with $f_{\rm env}=0.5$ and $0.6$ maintain radii comparable to or larger than the observed value of $9.3\,R_{\oplus}$, whereas models with $f_{\rm env}=0.3$ and $0.4$ have already contracted below this radius.
These results suggest that a sufficiently massive H/He envelope can sustain the highly inflated structure of Kepler-51 d over hundreds of Myr through secular cooling.

\subsubsection{Dependence on the total mass}
We consider three planetary masses, $M_p=4.5$, $5.6$, and $6.5\,M_{\oplus}$, spanning the observational uncertainty of Kepler-51 d, while fixing the equilibrium temperature at $T_{\rm eq}=350$ K and the envelope mass fraction at $f_{\rm env}=0.5$.

\begin{figure*}
{\includegraphics[width=0.99\textwidth]{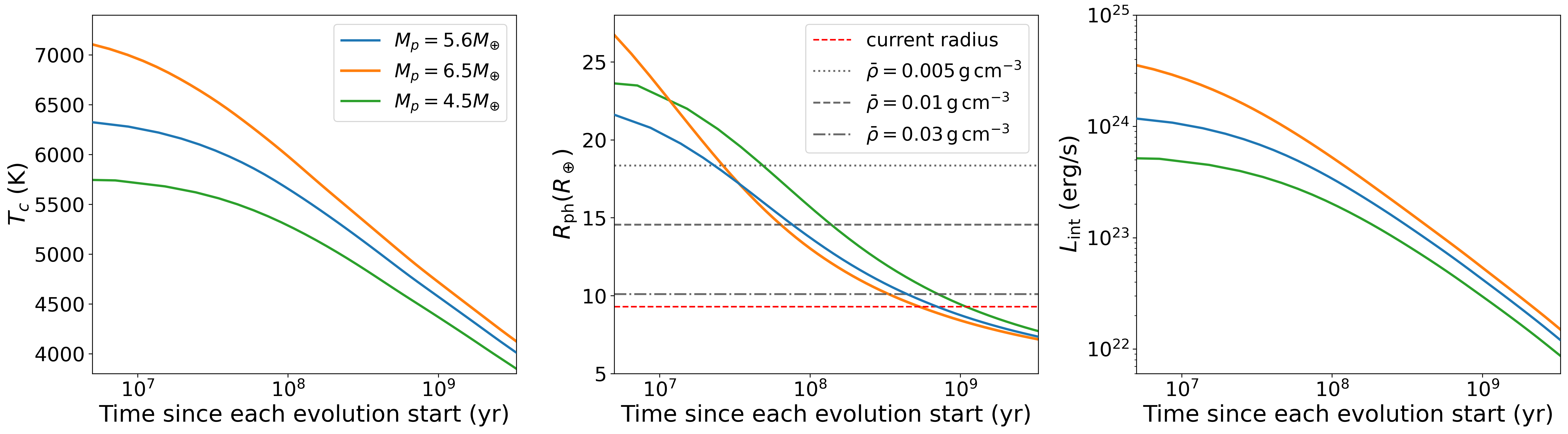}}
\caption{Same as Fig. \ref{fig_Teq}, but for different total planetary mass, with $T_{\rm eq}=350$ K and $f_{\rm env}=0.5$.}
\label{fig_mass}
\end{figure*}

The thermal evolution for different planetary masses is shown
in Fig.~\ref{fig_mass}. At a given $T_c$, more massive planets
have smaller radii due to their stronger self-gravity, resulting
in much denser planetary structures. At the same time, the magma ocean cools more slowly in more massive planets because, at a fixed envelope mass fraction, increasing planetary mass also increases envelope mass, leading to more thermal blanketing \citep{Hayashi1979,MatsuiAbe1986,Calder2026}.

Despite their slower core cooling, more massive planets remain less inflated at later stages of evolution, suggesting that stronger gravitational compression plays a more important role in determining the planetary radius. Nevertheless, all three models can maintain radii comparable to the observed radius of Kepler-51 d over its estimated age of $\sim500$ Myr.

\subsection{Dependence of Planetary Radius on Planetary Parameters}
\label{sec_parameter_space}
In Section \ref{sec_evo}, we have examined how the equilibrium temperature, envelope mass fraction, and total planetary mass individually affect the thermal evolution of super-puffs. For observed super-puffs, the planetary mass is often constrained by observations, and the equilibrium temperature can be estimated from stellar and orbital properties. In contrast, the H/He envelope mass fraction remains uncertain. It is therefore important to determine how much H/He is required for planets of different masses and equilibrium temperatures to maintain their large radii over Gyr timescales.

\begin{figure*}
{\includegraphics[width=0.99\textwidth]{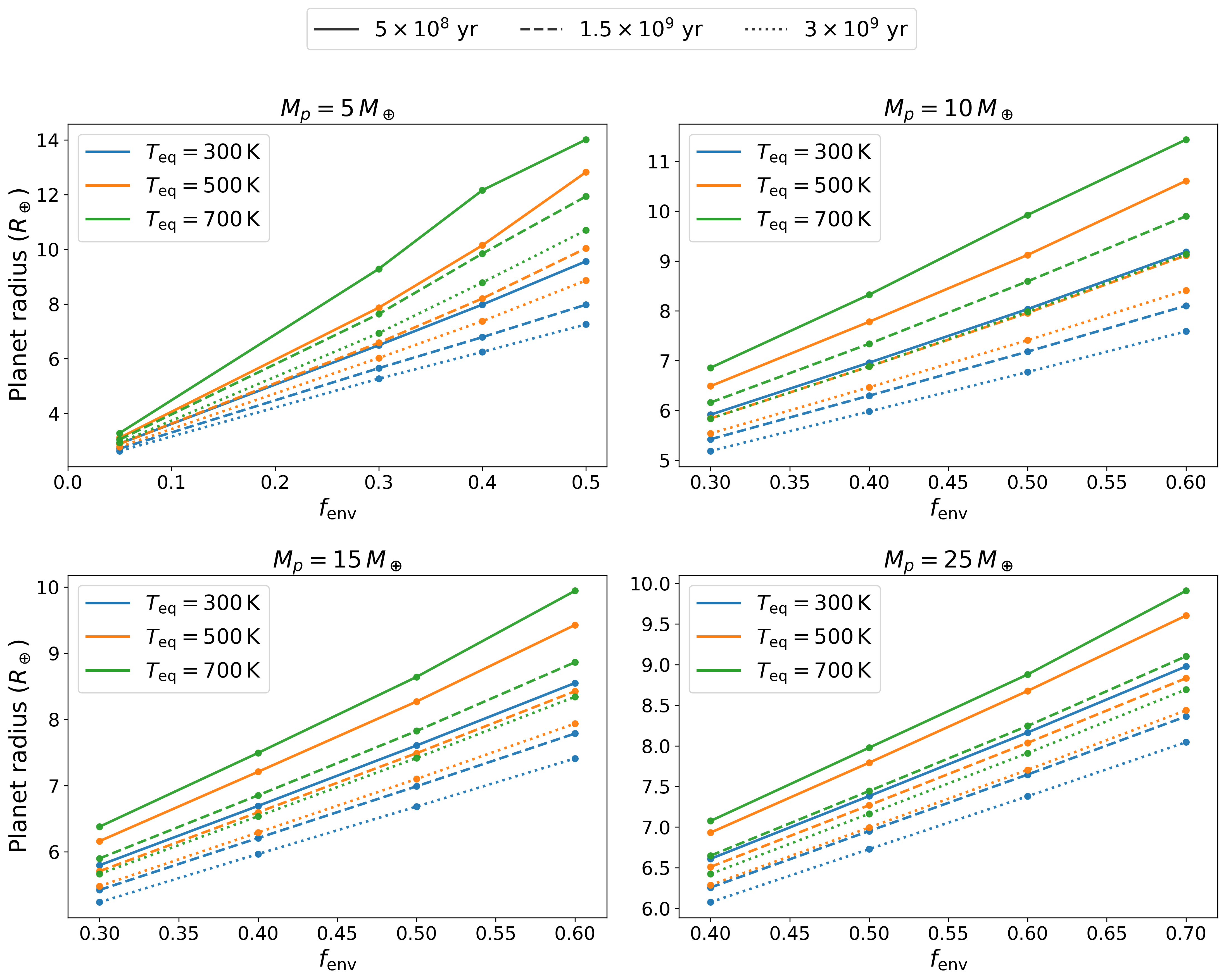}}
\caption{Planetary radius as a function of H/He envelope mass fraction $f_{\rm env}$ for different planetary masses, equilibrium temperatures, and cooling time. The upper-left, upper-right, bottom-left, and bottom-right panels correspond to $M_p=5$, $10$, $15$, and $25\,M_\oplus$, respectively. Different colors indicate equilibrium temperatures of $T_{\rm eq}=300$, $500$, and $700$ K, as labeled in each panel. Solid, dashed, and dotted lines represent the photospheric radii at ages of $5\times10^8$, $1.5\times10^9$, and $3\times10^9$ yr, respectively.}
\label{fig_parameter_space}
\end{figure*}

Fig.~\ref{fig_parameter_space} shows the planetary radius as a function of envelope mass fraction for four representative planetary masses, with different colors representing the equilibrium temperatures and different line styles representing the cooling time. At a given planetary mass and age, the radius increases strongly with $f_{\rm env}$, and higher equilibrium temperatures generally lead to larger radii. Planets gradually contract as they cool, and only those with sufficiently large envelope mass fractions can remain highly inflated for several Gyr.

The envelope mass fraction has a substantial effect on the planetary radius. For example, at $M_p=5\,M_\oplus$, planets with $f_{\rm env}=0.05$ have radii of only about $2.6$--$3\,R_\oplus$ after 3 Gyr, whereas those with $f_{\rm env}=0.5$ can retain radii of approximately $7$--$11\,R_\oplus$ at the same age. As planetary mass increases, stronger self-gravity tends to compress the gaseous envelope, requiring a larger $f_{\rm env}$ to maintain a comparable radius. Nevertheless, even planets with masses of several tens of Earth masses can remain substantially inflated if they possess sufficiently large H/He envelopes. These results suggest that a substantial envelope mass fraction is necessary for sustaining the exceptionally large radii of super-puffs over Gyr timescales.

The ability of large H/He envelopes to maintain such inflated structures can be attributed to two effects. First, the envelope provides strong thermal blanketing, which slows the cooling of the underlying rocky core. Second, the envelope itself contains a substantial reservoir of thermal and gravitational energy, which is gradually released through Kelvin--Helmholtz cooling and contraction. This energy reservoir allows the envelope to sustain its intrinsic luminosity over Gyr timescales while contracting gradually. Together, these effects enable super-puffs with substantial H/He envelopes to maintain large radii for several Gyr, even under relatively modest stellar irradiation.

\section{Discussion}\label{sec_discussion}
\subsection{The Role of Core Luminosity}\label{sec_Lc}
The intrinsic luminosity of a cooling planet arises from both the heat released by its rocky interior and the cooling and contraction of its H/He envelope. In our model, the core luminosity $L_c$ is determined by the secular cooling of the silicate magma ocean and provides the inner luminosity boundary condition for the envelope. The envelope also contributes to the intrinsic luminosity through its own thermal evolution, as described by $dL/dm=-T(\partial s/\partial t)_m$. As shown by the luminosity profiles in Fig. \ref{fig_profile}, the luminosity increases outward from the core–envelope boundary to the photosphere, indicating that the envelope itself releases a substantial amount of energy during cooling and contraction.

\begin{figure}
{\includegraphics[width=0.49\textwidth]{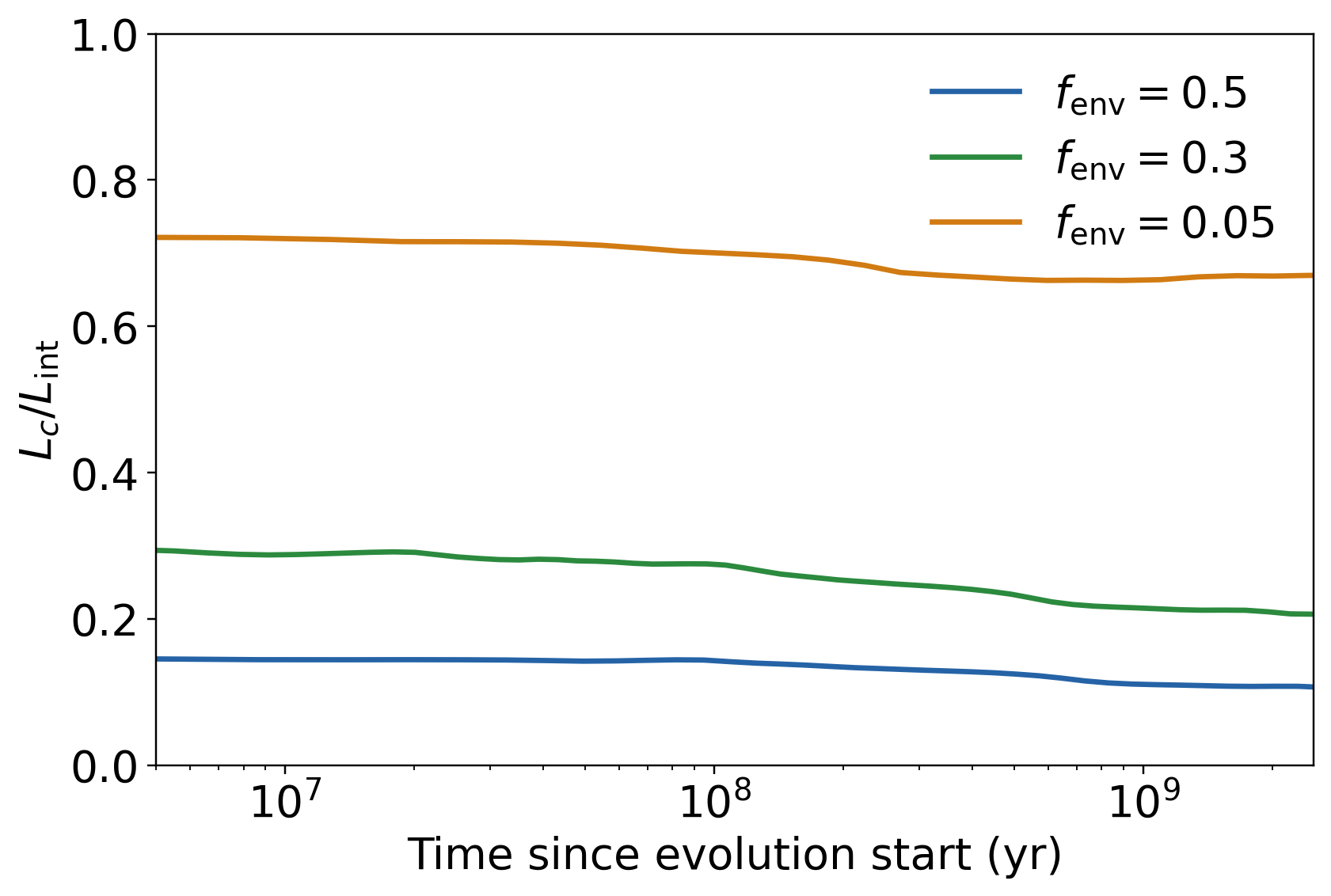}}
\caption{Evolution of the ratio of core luminosity to the photospheric intrinsic luminosity, $L_c/L_{\rm int}$, for planets with $M_p=5.6\,M_\oplus$, $T_{\rm eq}=350\,{\rm K}$, and envelope mass fractions $f_{\rm env}=0.5$, $0.3$, and $0.05$. The core contribution increases as the envelope mass fraction decreases.}
\label{fig_L}
\end{figure}

To quantify the relative contribution of core cooling, we calculate the ratio $L_c/L_{\rm int}$, where $L_{\rm int}$ is the intrinsic luminosity at the planetary photosphere. Fig. \ref{fig_L} shows its evolution for three envelope mass fractions. For $f_{\rm env}=0.5$, the core contributes only approximately $10$–$15\%$ of the total intrinsic luminosity, indicating that the envelope supplies most of the radiated energy. This fraction increases to approximately $20$–$30\%$ for $f_{\rm env}=0.3$ and $65$–$72\%$ for $f_{\rm env}=0.05$. In all three cases, the relative contribution of the core decreases only moderately over $\sim$ 1 Gyr. These results suggest that the envelope's own cooling and contraction dominate the luminosity of planets with massive H/He envelopes, whereas core cooling becomes increasingly important for planets with smaller envelope mass fractions.

\begin{figure*}
{\includegraphics[width=0.99\textwidth]{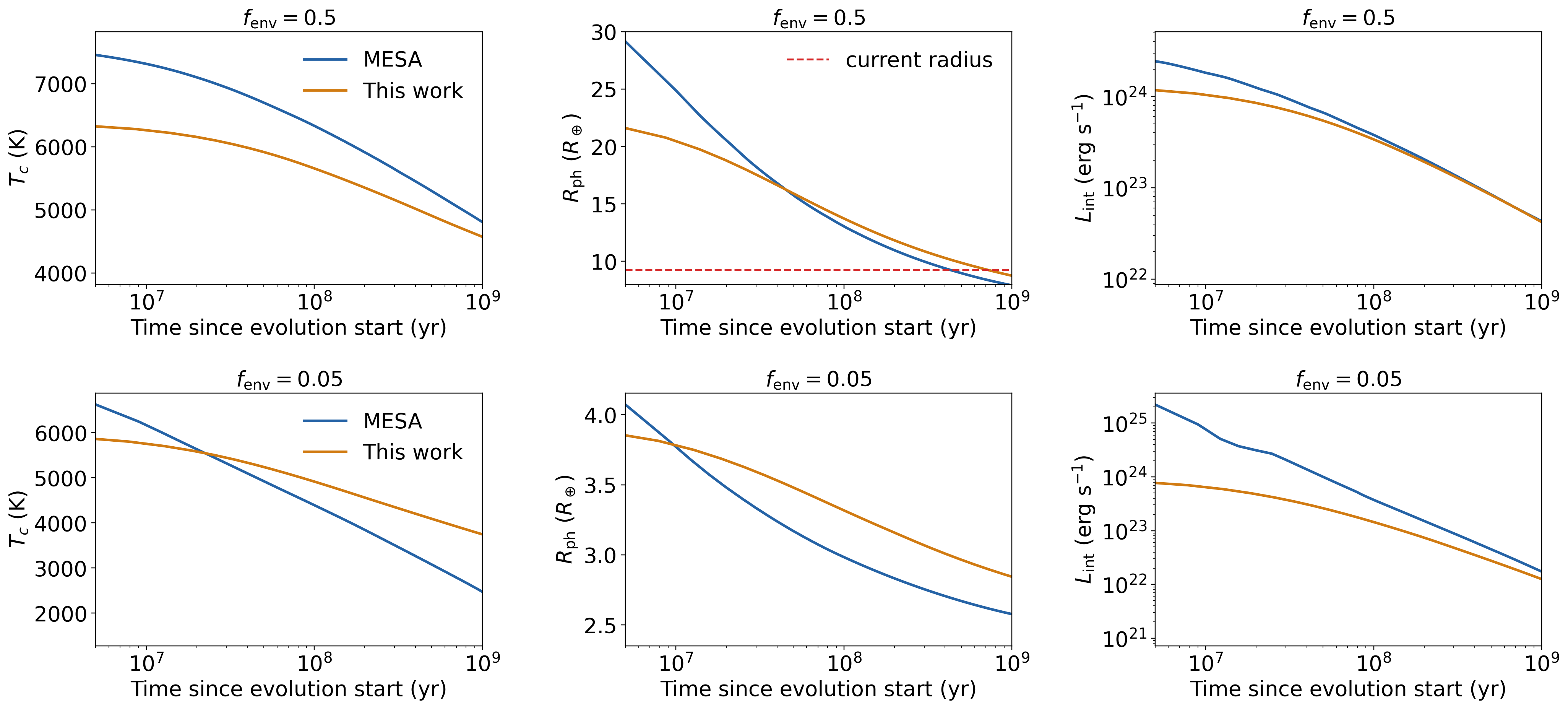}}
\caption{Comparison of the thermal evolution predicted by \texttt{MESA}
(blue) and our model (orange) for planets with $M_p=5.6\,M_\oplus$ and $T_{\rm eq}=350\,{\rm K}$. The upper and lower panels correspond to envelope mass fractions of $f_{\rm env}=0.5$ and $0.05$, respectively.
From left to right, the panels show the temperature $T_c$, photospheric
radius $R_{\rm ph}$, and photospheric intrinsic luminosity $L_{\rm int}$ as functions of time. Here $T_c$ denotes the magma-ocean temperature in our model, and it represents the temperature at the base of the H/He envelope in \texttt{MESA}, immediately above the inert core.
The horizontal red dashed line indicates the observed radius of Kepler-51 d ($9.32\,R_\oplus$). The \texttt{MESA} models adopt inert cores with $L_c=0$ during the cooling evolution, whereas our model includes
the secular cooling of the rocky interior.}
\label{fig_mesa}
\end{figure*}

We further compare our calculations with models computed
by Modules for Experiments in Stellar Astrophysics
(\texttt{MESA} r26.04.1; \citealt{Paxton2011,Paxton2013,
Paxton2015,Paxton2018,Paxton2019,Jermyn2023})
for $f_{\rm env}=0.5$ and $0.05$, as shown in Fig.~\ref{fig_mesa}. These \texttt{MESA} models have zero core luminosity during the cooling evolution, i.e., $L_c=L(m=M_c)=0$, but otherwise adopt the same planetary parameters as those described in Section \ref{sec_model}. To initiate the \texttt{MESA} calculation, we introduce an inert core with the prescribed mass and radius into a more massive, solar-composition gaseous model, and adjust the total planetary mass to achieve the desired envelope mass fraction; we add a constant core luminosity to heat the envelope and obtain an initial envelope profile. Once the model is thermally relaxed, we switch off the imposed core luminosity, reset the evolutionary clock to $t=0$, and follow the subsequent cooling and contraction for $1$ Gyr under the same irradiation. Both \texttt{MESA} and our model predict decreasing planetary radii and photospheric intrinsic luminosities over time. For $f_{\rm env}=0.5$, the two models approach similar radii and luminosities over the evolution. For $f_{\rm env}=0.05$, however, the \texttt{MESA} model undergoes more pronounced contraction, reaching a radius of approximately $2.6\,R_\oplus$ at $1$ Gyr, compared with approximately $2.9\,R_\oplus$ in our model. This difference is consistent with the greater relative importance of core cooling in planets with thin envelopes. 

Overall, these results indicate that core cooling and envelope cooling contribute differently to the evolution depending on the envelope mass fraction. For super-puffs with massive H/He envelopes, the envelope's thermal and gravitational energy reservoir plays a major role in sustaining the intrinsic luminosity during secular cooling. For planets with thinner envelopes, the core can supply a substantial fraction of the intrinsic luminosity and may have a correspondingly greater influence on the long-term radius evolution.

\subsection{Envelope Metallicity}
In our calculations presented in Section \ref{sec_results}, we adopt solar-metallicity opacities while neglecting the contribution of heavy elements to the EOS. This approximation is justified by the low mass fraction of heavy elements at solar composition ($Z\approx0.017$), which have a relatively small effect on the bulk thermodynamic properties of the envelope but can contribute substantially to its radiative opacity. However, the actual envelopes of super-puffs may be enriched in heavy elements, potentially affecting both their planetary structure and thermal evolution. At a given internal temperature, a higher mean molecular weight generally leads to a more compact envelope. On the other hand, heavy-element enrichment can increase the atmospheric opacity, reducing the loss of internal heat and slowing planetary contraction \citep{LopezFortney2014,Rogers2025}. The overall effect of metallicity on the planetary radius therefore depends on the competing influences of composition and opacity.

The envelope composition may also vary with depth. Chemical-equilibrium models suggest that interactions between a hot magma ocean and the overlying H/He envelope can enrich the deep atmosphere with rock-derived species \citep{SchlichtingYoung2022}. However, the efficiency of magma--atmosphere chemical exchange remains uncertain because compositional stratification near the magma-ocean surface may inhibit mixing and limit hydrogen exchange \citep{ModirroustaGalian2025}. If silicate vapor is present, the resulting mean molecular weight gradients can inhibit convection and modify the deep atmospheric temperature structure, thereby affecting the planetary radius and thermal evolution \citep{MisenerSchlichting2022}. Moreover, geological outgassing, another potential enrichment mechanism, occurs predominantly during magma-ocean crystallization because volatiles are highly incompatible in crystallized rock \citep{Lichtenberg2021,MiyazakiKorenaga2022}. We find that super-puffs maintain magma oceans for Gyr timescales, suggesting that outgassing is unlikely to contribute significantly to their envelope metallicities.

For Kepler-51d, recent JWST observations show no significant molecular absorption features, and the atmospheric models incorporating high-altitude hazes favor a relatively low metallicity of $\sim1$--$10$ times solar \citep{LibbyRoberts2026}. However, this inference is model-dependent, and the composition of the deeper envelope remains unconstrained by these observations. Consequently, the envelope mass fractions inferred from our models depend on the assumed composition, and accounting for heavy-element enrichment and compositional gradients could modify our quantitative results.

\subsection{Origin of Massive Envelopes}
Our results show that super-puffs can maintain their large radii over hundreds of millions to billions of years if they possess sufficiently massive H/He envelopes. However, the origin of such envelopes remains uncertain. As discussed by \citet{Marston2026}, super-puffs may provide useful constraints on core-accretion models. In some of our results, the envelope mass fraction approaches $f_{\rm env}=0.5$, corresponding to a gas-to-core mass ratio $M_{\rm env}/M_c\sim1$. In standard core-accretion models, this is approximately the regime in which the self-gravity of the envelope becomes important and gas accretion may enter the runaway phase \citep{Pollack1996,LeeChiang2016}. Explaining how low-mass planets acquire such large envelopes without growing into gas giants therefore requires understanding the conditions under which runaway begins and how gas accretion is eventually terminated.

Predictions for the onset of runaway gas accretion vary widely with assumptions about the structure and thermal evolution of the envelope. Early models find a wide range of possible critical core masses. Isothermal envelopes can yield $M_{\rm crit}\lesssim0.2\,M_\oplus$ \citep{Sasaki1989}, while radiative models give values of order $\sim10\,M_\oplus$ \citep{Stevenson1982}. In contrast, \citet{PerriCameron1974} obtained $M_{\rm crit}\sim115\,M_\oplus$ for an adiabatic envelope under their adopted nebular conditions. Models containing both radiative and convective regions generally predict critical core masses ranging from a few to several tens of Earth masses, depending on the opacity, orbital distance, and planetesimal accretion rate \citep{Mizuno1980,PapaloizouTerquem1999,Ikoma2000,Piso2015}. These estimates are based on different assumptions and, in some cases, different criteria for the onset of runaway. In particular, some studies determine the critical mass from the existence of hydrostatic solutions, while time-dependent models consider whether runaway gas accretion occurs within the lifetime of the protoplanetary disk. Thus, there is no universal critical core mass applicable to all formation environments.

The final envelope mass also depends on the time available for gas accretion and the supply of protoplanetary nebular gas. For example, \citet{LeeChiang2016} proposed that super-puffs could acquire substantial envelopes in cool, dust-free regions of protoplanetary disks, where efficient cooling allows rapid gas accretion. The subsequent dispersal of the disk could then limit further growth. Even after formation, atmospheric mass loss may reduce the envelope mass fraction, particularly during the early evolution of highly extended planets \citep{OwenWu2016,ModirroustaGalian2023,Tang2025Escape}. Our calculations do not account for either gas accretion or atmospheric escape, but instead examine the thermal evolution of planets that have already acquired and retained massive H/He envelopes.

 Our results suggest that massive envelopes can sustain highly inflated structures over Gyr timescales. Indeed, the wide range of critical core masses predicted by formation models suggests that super-puffs may provide useful constraints on the physical conditions governing gas accretion and its transition to runaway. Combining the observed properties of super-puffs with models of their formation and subsequent evolution could help determine which conditions allow low-mass cores to acquire and retain such large envelopes.

\section{Conclusion}\label{sec_conclusion}
We have investigated the long-term thermal evolution of super-puffs with massive H/He envelopes. Using a one-dimensional planetary structure and thermal evolution model, we follow the coupled cooling of the rocky core and the surrounding H/He envelope. Two processes play important roles in this evolution: (1) the H/He envelope acts as a thermal blanket, reducing heat loss from the underlying core magma ocean; and (2) the envelope releases energy through Kelvin–Helmholtz cooling and contraction, contributing to the planet’s intrinsic luminosity and governing the evolution of its entropy and radius. Our results show that planets with sufficiently massive H/He envelopes can maintain highly inflated radii over hundreds of millions to billions of years through secular cooling. For Kepler-51 d, $M_p=5.6\,M_\oplus$, $T_{\rm eq}=350$ K, and $f_{\rm env}=0.5$ produce a radius comparable to the observed $9.3\,R_\oplus$ at an age of approximately 500 Myr. Fig.~\ref{fig_parameter_space} summarizes our key findings for a range of $M_p$, $f_{\rm env}$, $T_{\rm eq}$ and ages; these can be used to constrain $f_{\rm env}$ for the observed super-puffs.

In this paper, we have examined how the evolution of the planetary radius depends on the equilibrium temperature, envelope mass fraction, and total planetary mass. Higher equilibrium temperatures warm the outer envelope, reducing the rate at which a planet loses internal heat and slowing its cooling and contraction. At fixed total mass, increasing the envelope mass fraction produces a more extended planetary structure, while core cooling scales only weakly with $f_{\rm env}$ at later stages. At fixed envelope mass fraction, more massive planets have smaller radii because of their stronger self-gravity, despite their slower core cooling. Our calculations over a broader parameter space further show that Jupiter-like radii can persist for Gyr timescales, provided that the H/He envelopes are sufficiently massive and the initial core temperatures are sufficiently high.

These results suggest that massive H/He envelopes offer a viable explanation for the large radii of super-puffs without invoking additional heating mechanisms or invoking high-altitude dust and hazes to artificially increase the apparent planetary radius. However, our calculations assume that these envelopes have already been accreted and retained, without accounting for atmospheric mass loss. How such massive envelopes can form and survive, particularly around low-mass rocky cores, remains an open question in understanding the origin of super-puffs.

\begin{acknowledgments}
X.H. thanks Yanqin Wu and Jun Yan Lau for useful discussions. X.H. is supported by the T.D. Lee scholarship. D.M.G. and R.W. acknowledge support from the National Natural Science Foundation of China (grant No. W2632045). 
This research has made use of the NASA Exoplanet Archive, which is operated by the California Institute of Technology, under contract with the National Aeronautics and Space Administration under the Exoplanet Exploration Program.

\end{acknowledgments}

\appendix
\section{Super-Puff CATALOG}
\label{ap1}
% This file is an AASTeX-compatible LaTeX table fragment.
% It uses longtable, booktabs, amsmath, and natbib; load these packages in your main manuscript.
% Produced from PS_2026-10-27_latest_per_planet_RETAIN_ALL.csv.
% All 68 catalog rows retained. Upper-limit rows and incomplete rows are NOT removed.
% Requires \bibliography{mybib-2,missing_planet_references} or equivalent.
\begingroup
\scriptsize
\setlength{\LTleft}{0pt}
\setlength{\LTright}{0pt}
\setlength{\tabcolsep}{3pt}
\renewcommand{\arraystretch}{1.13}
\begin{longtable}{@{}lccccc@{}}
\caption{Low-density planet catalog with planetary parameters and stellar ages. The data were retrieved from the NASA Exoplanet Archive on 2026 October 27 \citep{Christiansen_2025,NASAExoplanetArchive2026}.
The listed parameters are planetary radius $R_p$, planetary mass $M_p$, mean density $\rho$, equilibrium temperature $T_{\rm eq}$ (assuming zero Bond albedo), and host star age $t_\star$.}
\label{tab:superpuff_catalog}\\
\toprule
Planet & $R_{\rm p}\,(R_\oplus)$ & $M_{\rm p}\,(M_\oplus)$ & $\rho\,(\mathrm{g\,cm^{-3}})$ & $T_{\rm eq}\,(\mathrm{K})$ & $t_\star\,(\mathrm{Gyr})$ \\
\midrule
\endfirsthead
\multicolumn{6}{l}{\emph{Table \thetable{} (continued)}}\\
\toprule
Planet & $R_{\rm p}\,(R_\oplus)$ & $M_{\rm p}\,(M_\oplus)$ & $\rho\,(\mathrm{g\,cm^{-3}})$ & $T_{\rm eq}\,(\mathrm{K})$ & $t_\star\,(\mathrm{Gyr})$ \\
\midrule
\endhead
\midrule
\multicolumn{6}{r}{Continued on next page}\\
\endfoot
\bottomrule
\endlastfoot
HAT-P-18 b & $10.7^{+0.4}_{-0.3}$ & $55.5\pm1.7$ & $0.249^{+0.024}_{-0.026}$ & $841$ & $12.4^{+1.4}_{-6.4}$ \\
HATS-46 b & $10.7\pm0.3$ & $50\pm13$ & $0.227\pm0.065$ & $1082$ & $8.4\pm1.9$ \\
HATS-62 b & $11.83\pm0.28$ & $<56.9$ & $0.076\pm0.053$ & $1237$ & $9.55^{+0.99}_{-1.55}$ \\
HATS-8 b & $9.79^{+1.38}_{-0.84}$ & $43.9\pm6$ & $0.259\pm0.091$ & $1324$ & $5.1\pm1.7$ \\
HD 152843 c & $5.94^{+0.18}_{-0.16}$ & $9.7^{+2.0}_{-1.9}$ & $0.253^{+0.059}_{-0.057}$ & $945$ & $3.97\pm0.75$ \\
HIP 41378 f & $9.470\pm0.070$ & $25.0\pm5.0$ & $0.16^{+0.03}_{-0.04}$ & $294$ & $2.07^{+0.36}_{-0.27}$ \\
HIP 67522 b & $10.2\pm{1.8}$ & $13.8\pm1.0$ & $0.072\pm0.038$ & $1175$ & $0.0170\pm0.0020$ \\
K2-141 c & $7.0^{+4.6}_{-2.8}$ & $<8.00$ & $<0.081$ & $695$ & $6.3^{+6.6}_{-4.7}$ \\
K2-24 c & $7.93^{+0.12}_{-0.13}$ & $16.40^{+1.30}_{-0.20}$ & $0.181^{+0.017}_{-0.009}$ & $591$ & $4.9\pm1.7$ \\
KOI-351 g & $8.10\pm0.80$ & $15.0\pm1.3$ & $0.156\pm0.048$ & $386$ & $\ldots$ \\
Kepler-177 c & $8.73^{+0.36}_{-0.34}$ & $14.7^{+2.7}_{-2.5}$ & $0.121^{+0.027}_{-0.025}$ & $625$ & $4.37^{+3.63}_{-2.55}$ \\
Kepler-18 d & $6.98\pm0.33$ & $16.4\pm1.4$ & $0.27\pm0.03$ & $818$ & $10.72^{+2.00}_{-2.83}$ \\
Kepler-223 e & $4.60^{+0.27}_{-0.41}$ & $4.8^{+1.4}_{-1.2}$ & $0.28^{+0.12}_{-0.08}$ & $906$ & $4.27^{+3.92}_{-2.48}$ \\
Kepler-30 d & $8.8\pm0.5$ & $23.1\pm2.7$ & $0.19\pm0.02$ & $365$ & $2.0\pm0.8$ \\
Kepler-33 d & $4.6^{+1.0}_{-0.9}$ & $4.1^{+1.7}_{-2.0}$ & $0.2\pm0.2$ & $942$ & $4.27^{+0.74}_{-1.03}$ \\
Kepler-359 c & $4.80^{+1.00}_{-0.90}$ & $2.9^{+2.4}_{-1.9}$ & $0.1^{+0.2}_{-0.1}$ & $578$ & $3.02^{+1.36}_{-1.39}$ \\
Kepler-359 d & $4.60\pm0.90$ & $2.7^{+2.5}_{-1.5}$ & $0.1^{+0.2}_{-0.1}$ & $525$ & $3.02^{+1.36}_{-1.39}$ \\
Kepler-47 c & $4.65^{+0.09}_{-0.07}$ & $3.2^{+2.2}_{-1.2}$ & $0.170^{+0.09}_{-0.07}$ & $265$ & $\ldots$ \\
Kepler-51 b & $6.83\pm0.13$ & $3.5\pm2.1$ & $0.060\pm0.037$ & $543$ & $0.50\pm0.25$ \\
Kepler-51 c & $6.4\pm1.4$ & $5.65\pm0.18$ & $0.140\pm0.066$ & $439$ & $0.50\pm0.25$ \\
Kepler-51 d & $9.32\pm0.18$ & $5.6\pm1.2$ & $0.038\pm0.009$ & $350$ & $0.50\pm0.25$ \\
Kepler-79 d & $7.15\pm0.20$ & $5.3\pm0.9$ & $0.08\pm0.02$ & $642$ & $3.44^{+0.60}_{-0.91}$ \\
Kepler-87 c & $6.14\pm0.29$ & $6.40\pm0.80$ & $0.152\pm0.019$ & $403$ & $7.50\pm0.50$ \\
TOI-1135 b & $9.92\pm0.31$ & $<58.80$ & $0.150\pm0.09$ & $1171$ & $3.3^{+2.2}_{-1.6}$ \\
TOI-1173 b & $8.78^{+0.28}_{-0.26}$ & $28.3^{+1.0}_{-1.0}$ & $0.230^{+0.019}_{-0.026}$ & $945$ & $8.7^{+2.0}_{-1.9}$ \\
TOI-1420 b & $11.89\pm0.33$ & $25.1\pm3.8$ & $0.082\pm0.015$ & $957$ & $<10.7$ \\
TOI-1836 b & $8.28^{+0.20}_{-0.17}$ & $24.7^{+6.1}_{-5.9}$ & $0.240^{+0.07}_{-0.08}$ & $975$ & $6.6\pm1.8$ \\
TOI-2328 b & $10.0^{+0.3}_{-0.4}$ & $51\pm6$ & $0.29^{+0.04}_{-0.05}$ & $842$ & $4.0^{+2.3}_{-2.7}$ \\
TOI-2525 b & $8.68\pm0.11$ & $27\pm2$ & $0.225^{+0.015}_{-0.014}$ & $560$ & $4.0^{+4.3}_{-2.6}$ \\
TOI-3976 A b & $12.27^{+0.40}_{-0.39}$ & $56^{+12}_{-11}$ & $0.165^{+0.041}_{-0.037}$ & $1295$ & $3.8^{+1.5}_{-1.3}$ \\
TOI-4409 b & $7.44\pm0.19$ & $14.9^{+5.1}_{-5.4}$ & $0.200^{+0.069}_{-0.070}$ & $328$ & $7.0\pm2.0$ \\
TOI-4507 b & $8.20\pm0.10$ & $<20.00$ & $<0.200$ & $465$ & $0.70^{+0.80}_{-0.50}$ \\
TOI-5398 b & $10.30\pm0.40$ & $58.7^{+5.7}_{-5.6}$ & $0.29\pm0.05$ & $947$ & $0.65\pm0.15$ \\
TOI-5628 A b & $8.30\pm0.40$ & $29\pm13$ & $0.29\pm0.14$ & $492$ & $\ldots$ \\
TOI-7510 b & $7.31\pm0.23$ & $18.1\pm1.7$ & $0.255\pm0.026$ & $878$ & $4.4^{+1.7}_{-1.6}$ \\
TOI-791 b & $11.13\pm0.37$ & $9.5^{+1.6}_{-1.8}$ & $0.037^{+0.009}_{-0.007}$ & $475$ & $\ldots$ \\
TOI-791 c & $12.95\pm0.45$ & $18.6^{+2.2}_{-2.1}$ & $0.047^{+0.008}_{-0.007}$ & $397$ & $\ldots$ \\
V1298 Tau b & $9.41\pm0.57$ & $13.1\pm5.3$ & $0.09\pm0.04$ & $666$ & $0.02\pm0.01$ \\
V1298 Tau c & $5.08\pm0.37$ & $4.7\pm0.6$ & $0.20\pm0.05$ & $953$ & $0.02\pm0.01$ \\
V1298 Tau d & $6.53\pm0.42$ & $6.0\pm0.7$ & $0.12\pm0.03$ & $831$ & $0.02\pm0.01$ \\
V1298 Tau e & $10.17\pm0.75$ & $15.3\pm4.2$ & $0.08\pm0.03$ & $527$ & $0.02\pm0.01$ \\
WASP-107 b & $10.5\pm0.3$ & $33.0^{+1.2}_{-1.3}$ & $0.157^{+0.012}_{-0.014}$ & $738$ & $3.40\pm0.70$ \\
WASP-127 b & $14.69^{+0.28}_{-0.33}$ & $52.4^{+6.8}_{-5.5}$ & $0.097^{+0.013}_{-0.012}$ & $1404$ & $11.4\pm1.8$ \\
WASP-193 b & $14.78^{+0.63}_{-0.54}$ & $35.6^{+9.2}_{-10.8}$ & $0.060\pm0.019$ & $1250$ & $6.8^{+3.1}_{-2.4}$ \\
WASP-195 b & $10\pm1$ & $33.1^{+9.5}_{-9.9}$ & $0.16\pm0.06$ & $1522$ & $0.60\pm0.20$ \\
\end{longtable}
\endgroup

\bibliography{mybib}{}

@article{Chabrier2021,
doi = {10.3847/1538-4357/abfc48},
url = {https://doi.org/10.3847/1538-4357/abfc48},
year = {2021},
month = {aug},
publisher = {The American Astronomical Society},
volume = {917},
number = {1},
pages = {4},
author = {Chabrier, Gilles and Debras, Florian},
title = {A New Equation of State for Dense Hydrogen–Helium Mixtures. II. Taking into Account Hydrogen–Helium Interactions},
journal = {The Astrophysical Journal}
}

@article{JontofHutter2014,
  doi       = {10.1088/0004-637X/785/1/15},
  url       = {https://doi.org/10.1088/0004-637X/785/1/15},
  year      = {2014},
  month     = {mar},
  publisher = {The American Astronomical Society},
  volume    = {785},
  number    = {1},
  pages     = {15},
  author    = {Jontof-Hutter, Daniel and Lissauer, Jack J. and Rowe, Jason F. and Fabrycky, Daniel C.},
  title     = {{Kepler-79}'s Low Density Planets},
  journal   = {The Astrophysical Journal},
}

@article{Masuda2014,
  doi       = {10.1088/0004-637X/783/1/53},
  url       = {https://doi.org/10.1088/0004-637X/783/1/53},
  year      = {2014},
  month     = {feb},
  publisher = {The American Astronomical Society},
  volume    = {783},
  number    = {1},
  pages     = {53},
  author    = {Masuda, Kento},
  title     = {Very Low Density Planets around {Kepler-51} Revealed with Transit Timing Variations and an Anomaly Similar to a Planet--Planet Eclipse Event},
  journal   = {The Astrophysical Journal},
}

@article{Barkaoui2024,
  doi       = {10.1038/s41550-024-02259-y},
  url       = {https://doi.org/10.1038/s41550-024-02259-y},
  year      = {2024},
  month     = {may},
  publisher = {Springer Nature},
  volume    = {8},
  number    = {7},
  pages     = {909--919},
  author    = {Barkaoui, Khalid and Pozuelos, Francisco J. and Hellier, Coel and Smalley, Barry and Nielsen, Louise D. and Niraula, Prajwal and Gillon, Micha{\"e}l and de Wit, Julien and M{\"u}ller, Simon and Dorn, Caroline and Helled, Ravit and Jehin, Emmanuel and Demory, Brice-Olivier and Van Grootel, Valerie and Soubkiou, Abderahmane and Ghachoui, Mourad and Anderson, David R. and Benkhaldoun, Zouhair and Bouchy, Francois and Burdanov, Artem and Delrez, Laetitia and Ducrot, Elsa and Garcia, Lionel and Jabiri, Abdelhadi and Lendl, Monika and Maxted, Pierre F. L. and Murray, Catriona A. and Pedersen, Peter Pihlmann and Queloz, Didier and Sebastian, Daniel and Turner, Oliver and Udry, Stephane and Timmermans, Mathilde and Triaud, Amaury H. M. J. and West, Richard G.},
  title     = {An extended low-density atmosphere around the Jupiter-sized planet {WASP-193 b}},
  journal   = {Nature Astronomy},
}

@article{Masuda2024,
  doi       = {10.3847/1538-3881/ad83d3},
  url       = {https://doi.org/10.3847/1538-3881/ad83d3},
  year      = {2024},
  month     = {dec},
  publisher = {The American Astronomical Society},
  volume    = {168},
  number    = {6},
  pages     = {294},
  author    = {Masuda, Kento and Libby-Roberts, Jessica E. and Livingston, John H. and Stevenson, Kevin B. and Gao, Peter and Vissapragada, Shreyas and Fu, Guangwei and Han, Te and Greklek-McKeon, Michael and Mahadevan, Suvrath and Agol, Eric and Bello-Arufe, Aaron and Berta-Thompson, Zachory K. and Ca{\~n}as, Caleb I. and Chachan, Yayaati and Hebb, Leslie and Hu, Renyu and Kawashima, Yui and Knutson, Heather A. and Morley, Caroline V. and Murray, Catriona A. and Ohno, Kazumasa and Tokadjian, Armen and Zhang, Xi and Welbanks, Luis and Nixon, Matthew C. and Freedman, Richard and Narita, Norio and Fukui, Akihiko and de Leon, Jerome P. and Mori, Mayuko and Pall{\'e}, Enric and Murgas, Felipe and Parviainen, Hannu and Esparza-Borges, Emma and Jontof-Hutter, Daniel and Collins, Karen A. and Benni, Paul and Barkaoui, Khalid and Pozuelos, Francisco J. and Gillon, Micha{\"e}l and Jehin, Emmanu{\"e}l and Benkhaldoun, Zouhair and Hawley, Suzanne and Lin, Andrea S. J. and Stef{\'a}nsson, Gu{\dh}mundur and Bieryla, Allyson and Yilmaz, Mesut and Senavci, Hakan Volkan and Girardin, Eric and Marino, Giuseppe and Wang, Gavin},
  title     = {A Fourth Planet in the {Kepler-51} System Revealed by Transit Timing Variations},
  journal   = {The Astronomical Journal},
}

@article{LibbyRoberts2026,
  doi       = {10.3847/1538-3881/ae33c0},
  url       = {https://doi.org/10.3847/1538-3881/ae33c0},
  year      = {2026},
  month     = {mar},
  publisher = {The American Astronomical Society},
  volume    = {171},
  number    = {4},
  pages     = {221},
  author    = {Libby-Roberts, Jessica E. and Bello-Arufe, Aaron and Berta-Thompson, Zachory K. and Ca{\~n}as, Caleb I. and Chachan, Yayaati and Hu, Renyu and Kawashima, Yui and Murray, Catriona and Ohno, Kazumasa and Tokadjian, Armen and Mahadevan, Suvrath and Masuda, Kento and Hebb, Leslie and Morley, Caroline and Fu, Guangwei and Gao, Peter and Stevenson, Kevin B.},
  title     = {The James Webb Space Telescope {NIRSpec-PRISM} Transmission Spectrum of the Super-puff, {Kepler-51d}},
  journal   = {The Astronomical Journal},
}

@article{Dransfield2026,
  doi       = {10.1093/mnras/stag864},
  url       = {https://doi.org/10.1093/mnras/stag864},
  year      = {2026},
  month     = {jul},
  publisher = {Oxford University Press},
  volume    = {549},
  number    = {4},
  pages     = {stag864},
  author    = {Dransfield, Georgina and Petit, Antoine C. and Triaud, Amaury H. M. J. and Guillot, Tristan and Schmider, Fran{\c{c}}ois-Xavier and Abe, Lyu and Agabi, Abdelkrim and Barkaoui, Khalid and Baycroft, Thomas A. and Bendjoya, Philippe and Brahm, Rafael and Collins, Karen A. and Edwards, Billy and Evans, Phil and Freckelton, Alix V. and Grieves, Nolan and Howell, Steve B. and Mallia, Franco and Mekarnia, Djamel and Psaridi, Angelica and Sebastian, Daniel and Stassun, Keivan G. and Stockdale, Chris and Stokholm, Amalie and Suarez, Olga and Tan, Thiam-Guan and Timmermans, Mathilde and Watkins, Cristilyn N. and Ziegler, Carl and Soubkiou, Abderahmane and Bouchy, Fran{\c{c}}ois and Cointepas, Marion and Deloupy, Vincent and G{\"u}nther, Maximilian N. and Gillon, Micha{\"e}l and Isopi, Giovanni and Jehin, Emmanuel and Jenkins, Jon M. and Jord{\'a}n, Andr{\'e}s and Nielsen, Martin B. and Seager, Sara and Shporer, Avi and Seidel, Julia V. and Steiner, Michal and Trifonov, Trifon and Twicken, Joseph D. and Winn, Joshua N. and Zapparata, Aldo},
  title     = {{ASTEP} confirmation of a pair of long-period Jupiter-sized planets with extremely low densities transiting {TOI-791}},
  journal   = {Monthly Notices of the Royal Astronomical Society},
}

@article{Lund2019,
  doi       = {10.3847/1538-3881/ab5280},
  url       = {https://doi.org/10.3847/1538-3881/ab5280},
  year      = {2019},
  month     = {dec},
  publisher = {The American Astronomical Society},
  volume    = {158},
  number    = {6},
  pages     = {248},
  author    = {Lund, Mikkel N. and Knudstrup, Emil and Silva Aguirre, V{\'i}ctor and Basu, Sarbani and Chontos, Ashley and von Essen, Carolina and Chaplin, William J. and Bieryla, Allyson and Casagrande, Luca and Vanderburg, Andrew and Huber, Daniel and Kane, Stephen R. and Albrecht, Simon and Latham, David W. and Davies, Guy R. and Becker, Juliette C. and Rodriguez, Joseph E.},
  title     = {Asteroseismology of the Multiplanet System {K2-93}},
  journal   = {The Astronomical Journal},
}

@article{LopezFortney2014,
  doi       = {10.1088/0004-637X/792/1/1},
  url       = {https://doi.org/10.1088/0004-637X/792/1/1},
  year      = {2014},
  month     = {sep},
  publisher = {The American Astronomical Society},
  volume    = {792},
  number    = {1},
  pages     = {1},
  author    = {Lopez, Eric D. and Fortney, Jonathan J.},
  title     = {Understanding the Mass-Radius Relation for Sub-Neptunes: Radius as a Proxy for Composition},
  journal   = {The Astrophysical Journal},
}

@article{LibbyRoberts2020,
  doi       = {10.3847/1538-3881/ab5d36},
  url       = {https://doi.org/10.3847/1538-3881/ab5d36},
  year      = {2020},
  month     = {jan},
  publisher = {The American Astronomical Society},
  volume    = {159},
  number    = {2},
  pages     = {57},
  author    = {Libby-Roberts, Jessica E. and Berta-Thompson, Zachory K. and D{\'e}sert, Jean-Michel and Masuda, Kento and Morley, Caroline V. and Lopez, Eric D. and Deck, Katherine M. and Fabrycky, Daniel and Fortney, Jonathan J. and Line, Michael R. and Sanchis-Ojeda, Roberto and Winn, Joshua N.},
  title     = {The Featureless Transmission Spectra of Two Super-puff Planets},
  journal   = {The Astronomical Journal},
}

@article{Marston2026,
  doi       = {10.3847/1538-4357/ae78d1},
  url       = {https://doi.org/10.3847/1538-4357/ae78d1},
  year      = {2026},
  month     = {jul},
  publisher = {The American Astronomical Society},
  volume    = {1005},
  number    = {2},
  pages     = {204},
  author    = {Marston, Nicholas T. and Becker, Juliette and Howe, Alex R.},
  title     = {Do Super-puffs Defy Core Accretion? Population-wide Interior Structure Constraints},
  journal   = {The Astrophysical Journal},
}

@article{Wu2019,
  doi       = {10.3847/1538-4357/ab06f8},
  url       = {https://doi.org/10.3847/1538-4357/ab06f8},
  year      = {2019},
  month     = {mar},
  publisher = {The American Astronomical Society},
  volume    = {874},
  number    = {1},
  pages     = {91},
  author    = {Wu, Yanqin},
  title     = {Mass and Mass Scalings of Super-Earths},
  journal   = {The Astrophysical Journal},
}

@article{LeeChiang2016,
  doi       = {10.3847/0004-637X/817/2/90},
  url       = {https://doi.org/10.3847/0004-637X/817/2/90},
  year      = {2016},
  month     = {jan},
  publisher = {The American Astronomical Society},
  volume    = {817},
  number    = {2},
  pages     = {90},
  author    = {Lee, Eve J. and Chiang, Eugene},
  title     = {Breeding Super-Earths and Birthing Super-Puffs in Transitional Disks},
  journal   = {The Astrophysical Journal},
}

@article{Chachan2021,
  doi       = {10.3847/1538-4357/ac0bb6},
  url       = {https://doi.org/10.3847/1538-4357/ac0bb6},
  year      = {2021},
  month     = {sep},
  publisher = {The American Astronomical Society},
  volume    = {919},
  number    = {1},
  pages     = {63},
  author    = {Chachan, Yayaati and Lee, Eve J. and Knutson, Heather A.},
  title     = {Radial Gradients in Dust-to-gas Ratio Lead to Preferred Region for Giant Planet Formation},
  journal   = {The Astrophysical Journal},
}

@article{OwenWu2016,
  doi       = {10.3847/0004-637X/817/2/107},
  url       = {https://doi.org/10.3847/0004-637X/817/2/107},
  year      = {2016},
  month     = {jan},
  publisher = {The American Astronomical Society},
  volume    = {817},
  number    = {2},
  pages     = {107},
  author    = {Owen, James E. and Wu, Yanqin},
  title     = {Atmospheres of Low-mass Planets: The ``Boil-off''},
  journal   = {The Astrophysical Journal},
}

@article{Tang2025Structure,
  doi       = {10.3847/1538-4357/ade7ff},
  url       = {https://doi.org/10.3847/1538-4357/ade7ff},
  year      = {2025},
  month     = {aug},
  publisher = {The American Astronomical Society},
  volume    = {989},
  number    = {1},
  pages     = {28},
  author    = {Tang, Yao and Fortney, Jonathan J. and Nimmo, Francis and Thorngren, Daniel and Ohno, Kazumasa and Murray-Clay, Ruth},
  title     = {Reassessing Sub-Neptune Structure, Radii, and Thermal Evolution},
  journal   = {The Astrophysical Journal},
}

@article{Tang2025Escape,
  doi       = {10.3847/1538-4357/ae147a},
  url       = {https://doi.org/10.3847/1538-4357/ae147a},
  year      = {2025},
  month     = {dec},
  publisher = {The American Astronomical Society},
  volume    = {995},
  number    = {1},
  pages     = {20},
  author    = {Tang, Yao and Fortney, Jonathan J. and Murray-Clay, Ruth and Broome, Madelyn},
  title     = {Understanding the Origins of Super-puff Planets: A New Mass-loss Regime Coupled to Planetary Evolution},
  journal   = {The Astrophysical Journal},
}

@article{Millholland2019,
  doi       = {10.3847/1538-4357/ab4c3f},
  url       = {https://doi.org/10.3847/1538-4357/ab4c3f},
  year      = {2019},
  month     = {nov},
  publisher = {The American Astronomical Society},
  volume    = {886},
  number    = {1},
  pages     = {72},
  author    = {Millholland, Sarah C.},
  title     = {Tidally Induced Radius Inflation of Sub-Neptunes},
  journal   = {The Astrophysical Journal},
}

@article{WangDai2019,
  doi       = {10.3847/2041-8213/ab0653},
  url       = {https://doi.org/10.3847/2041-8213/ab0653},
  year      = {2019},
  month     = {feb},
  publisher = {The American Astronomical Society},
  volume    = {873},
  number    = {1},
  pages     = {L1},
  author    = {Wang, Lile and Dai, Fei},
  title     = {Dusty Outflows in Planetary Atmospheres: Understanding ``Super-puffs'' and Transmission Spectra of Sub-Neptunes},
  journal   = {The Astrophysical Journal Letters},
}

@article{GaoZhang2020,
  doi       = {10.3847/1538-4357/ab6a9b},
  url       = {https://doi.org/10.3847/1538-4357/ab6a9b},
  year      = {2020},
  month     = {feb},
  publisher = {The American Astronomical Society},
  volume    = {890},
  number    = {2},
  pages     = {93},
  author    = {Gao, Peter and Zhang, Xi},
  title     = {Deflating Super-puffs: Impact of Photochemical Hazes on the Observed Mass--Radius Relationship of Low-mass Planets},
  journal   = {The Astrophysical Journal},
}

@article{OhnoTanaka2021,
  doi       = {10.3847/1538-4357/ac1516},
  url       = {https://doi.org/10.3847/1538-4357/ac1516},
  year      = {2021},
  month     = {oct},
  publisher = {The American Astronomical Society},
  volume    = {920},
  number    = {2},
  pages     = {124},
  author    = {Ohno, Kazumasa and Tanaka, Yuki A.},
  title     = {Grain Growth in Escaping Atmospheres: Implications for the Radius Inflation of Super-Puffs},
  journal   = {The Astrophysical Journal},
}

@article{PiroVissapragada2020,
  doi       = {10.3847/1538-3881/ab7192},
  url       = {https://doi.org/10.3847/1538-3881/ab7192},
  year      = {2020},
  month     = {feb},
  publisher = {The American Astronomical Society},
  volume    = {159},
  number    = {4},
  pages     = {131},
  author    = {Piro, Anthony L. and Vissapragada, Shreyas},
  title     = {Exploring Whether Super-puffs Can Be Explained as Ringed Exoplanets},
  journal   = {The Astronomical Journal},
}

@article{Akinsanmi2020,
  doi       = {10.1051/0004-6361/202037618},
  url       = {https://doi.org/10.1051/0004-6361/202037618},
  year      = {2020},
  month     = {mar},
  publisher = {EDP Sciences},
  volume    = {635},
  pages     = {L8},
  author    = {Akinsanmi, B. and Santos, N. C. and Faria, J. P. and Oshagh, M. and Barros, S. C. C. and Santerne, A. and Charnoz, S.},
  title     = {Can Planetary Rings Explain the Extremely Low Density of {HIP 41378 f}?},
  journal   = {Astronomy and Astrophysics},
}

@article{Lu2025,
  doi       = {10.3847/1538-4357/ada4b2},
  url       = {https://doi.org/10.3847/1538-4357/ada4b2},
  year      = {2025},
  month     = {feb},
  publisher = {The American Astronomical Society},
  volume    = {980},
  number    = {1},
  pages     = {39},
  author    = {Lu, Tiger and Li, Gongjie and Cassese, Ben and Lin, D. N. C.},
  title     = {The Dynamical History of {HIP-41378 f}---Oblique Exorings Masquerading as a Puffy Planet},
  journal   = {The Astrophysical Journal},
}

@article{Umetani2025,
  doi       = {10.1093/pasj/psae101},
  url       = {https://doi.org/10.1093/pasj/psae101},
  year      = {2025},
  month     = {feb},
  publisher = {Oxford University Press},
  volume    = {77},
  number    = {1},
  pages     = {86--104},
  author    = {Umetani, Tsubasa and Aizawa, Masataka and Ezoe, Yuichiro and Ishisaki, Yoshitaka},
  title     = {Search for Exoplanetary Ring Systems with {TESS}},
  journal   = {Publications of the Astronomical Society of Japan},
}

@article{Ikoma2000,
  doi = {10.1086/309050},
  url = {https://doi.org/10.1086/309050},
  year = {2000},
  month = {jul},
  publisher = {The American Astronomical Society},
  volume = {537},
  number = {2},
  pages = {1013--1025},
  author = {Ikoma, Masahiro and Nakazawa, Kiyoshi and Emori, Hiroyuki},
  title = {Formation of Giant Planets: Dependences on Core Accretion Rate and Grain Opacity},
  journal = {The Astrophysical Journal},
}

@article{Mizuno1980,
  doi = {10.1143/PTP.64.544},
  url = {https://doi.org/10.1143/PTP.64.544},
  year = {1980},
  month = {aug},
  volume = {64},
  number = {2},
  pages = {544--557},
  author = {Mizuno, Hiroshi},
  title = {Formation of the Giant Planets},
  journal = {Progress of Theoretical Physics},
}

@article{PapaloizouTerquem1999,
  doi = {10.1086/307581},
  url = {https://doi.org/10.1086/307581},
  year = {1999},
  month = {aug},
  volume = {521},
  number = {2},
  pages = {823--838},
  author = {Papaloizou, John C. B. and Terquem, Caroline},
  title = {Critical Protoplanetary Core Masses in Protoplanetary Disks and the Formation of Short-Period Giant Planets},
  journal = {The Astrophysical Journal},
}

@article{PerriCameron1974,
  doi = {10.1016/0019-1035(74)90074-8},
  url = {https://doi.org/10.1016/0019-1035(74)90074-8},
  year = {1974},
  month = {aug},
  publisher = {Elsevier},
  volume = {22},
  number = {4},
  pages = {416--425},
  author = {Perri, F. and Cameron, A. G. W.},
  title = {Hydrodynamic Instability of the Solar Nebula in the Presence of a Planetary Core},
  journal = {Icarus},
}

@article{Piso2015,
  doi = {10.1088/0004-637X/800/2/82},
  url = {https://doi.org/10.1088/0004-637X/800/2/82},
  year = {2015},
  month = {feb},
  publisher = {The American Astronomical Society},
  volume = {800},
  number = {2},
  pages = {82},
  author = {Piso, Ana-Maria A. and Youdin, Andrew N. and Murray-Clay, Ruth A.},
  title = {Minimum Core Masses for Giant Planet Formation with Realistic Equations of State and Opacities},
  journal = {The Astrophysical Journal},
}

@article{Sasaki1989,
  year = {1989},
  month = {may},
  volume = {215},
  pages = {177--180},
  author = {Sasaki, S.},
  title = {Minimum Planetary Size for Forming Outer Jovian-type Planets---Stability of an Isothermal Atmosphere Surrounding a Protoplanet},
  journal = {Astronomy and Astrophysics},
}

@article{Stevenson1982,
  doi = {10.1016/0032-0633(82)90108-8},
  url = {https://doi.org/10.1016/0032-0633(82)90108-8},
  year = {1982},
  month = {aug},
  publisher = {Pergamon Press},
  volume = {30},
  number = {8},
  pages = {755--764},
  author = {Stevenson, D. J.},
  title = {Formation of the Giant Planets},
  journal = {Planetary and Space Science},
}

@article{Rogers2025,
  doi = {10.1093/mnras/staf628},
  url = {https://doi.org/10.1093/mnras/staf628},
  year = {2025},
  month = {may},
  publisher = {Oxford University Press},
  volume = {539},
  number = {3},
  pages = {2230--2241},
  author = {Rogers, James G.},
  title = {On the Road to the Radius Valley: Distinguishing between Gas Dwarfs and Water Worlds with Young Transiting Exoplanets},
  journal = {Monthly Notices of the Royal Astronomical Society},
}

@article{SchlichtingYoung2022,
  doi = {10.3847/PSJ/ac68e6},
  url = {https://doi.org/10.3847/PSJ/ac68e6},
  year = {2022},
  month = {may},
  publisher = {The American Astronomical Society},
  volume = {3},
  number = {5},
  pages = {127},
  author = {Schlichting, Hilke E. and Young, Edward D.},
  title = {Chemical Equilibrium between Cores, Mantles, and Atmospheres of Super-Earths and Sub-Neptunes and Implications for Their Compositions, Interiors, and Evolution},
  journal = {The Planetary Science Journal},
}

@article{MisenerSchlichting2022,
  doi = {10.1093/mnras/stac1732},
  url = {https://doi.org/10.1093/mnras/stac1732},
  year = {2022},
  month = {aug},
  publisher = {Oxford University Press},
  volume = {514},
  number = {4},
  pages = {6025--6037},
  author = {Misener, William and Schlichting, Hilke E.},
  title = {The Importance of Silicate Vapour in Determining the Structure, Radii, and Envelope Mass Fractions of Sub-Neptunes},
  journal = {Monthly Notices of the Royal Astronomical Society},
}

@article{Christiansen_2025,
doi = {10.3847/PSJ/ade3c2},
url = {https://doi.org/10.3847/PSJ/ade3c2},
year = {2025},
month = {aug},
publisher = {The American Astronomical Society},
volume = {6},
number = {8},
pages = {186},
author = {Christiansen, Jessie L. and McElroy, Douglas L. and Harbut, Marcy and Ciardi, David R. and Crane, Megan and Good, John and Hardegree-Ullman, Kevin K. and Kesseli, Aurora Y. and Lund, Michael B. and Lynn, Meca and Muthiar, Ananda and Nilsson, Ricky and Oluyide, Toba and Papin, Michael and Rivera, Amalia and Swain, Melanie and Susemiehl, Nicholas D. and Tam, Raymond and van Eyken, Julian and Beichman, Charles},
title = {The NASA Exoplanet Archive and Exoplanet Follow-up Observing Program: Data, Tools, and Usage},
journal = {The Planetary Science Journal},}

@article{Pollack1996,
  doi = {10.1006/icar.1996.0190},
  url = {https://doi.org/10.1006/icar.1996.0190},
  year = {1996},
  month = {nov},
  publisher = {Academic Press},
  volume = {124},
  number = {1},
  pages = {62--85},
  author = {Pollack, James B. and Hubickyj, Olenka and Bodenheimer, Peter and Lissauer, Jack J. and Podolak, Morris and Greenzweig, Yuval},
  title = {Formation of the Giant Planets by Concurrent Accretion of Solids and Gas},
  journal = {Icarus},
}

@ARTICLE{Guillot2010,
    author = {{Guillot}, Tristan},
    title = "{On the radiative equilibrium of irradiated planetary atmospheres}",
    journal = {Astronomy \& Astrophysics},
    year = {2010},
    month = sep,
    volume = {520},
    eid = {A27},
    pages = {A27},
    numpages = {13},
    doi = {10.1051/0004-6361/200913396},
    adsurl = {https://ui.adsabs.harvard.edu/abs/2010A%26A...520A..27G}
}

@ARTICLE{Seager2007,
    author = {{Seager}, S. and {Kuchner}, M. and
              {Hier-Majumder}, C. A. and {Militzer}, B.},
    title = "{Mass-Radius Relationships for Solid Exoplanets}",
    journal = {The Astrophysical Journal},
    year = {2007},
    volume = {669},
    number = {2},
    pages = {1279--1297},
    doi = {10.1086/521346}
}

@ARTICLE{CowanAgol2011,
    author = {{Cowan}, Nicolas B. and {Agol}, Eric},
    title = "{The Statistics of Albedo and Heat Recirculation on Hot Exoplanets}",
    journal = {The Astrophysical Journal},
    year = {2011},
    month = feb,
    volume = {729},
    number = {1},
    pages = {54},
    doi = {10.1088/0004-637X/729/1/54},
    adsurl = {https://ui.adsabs.harvard.edu/abs/2011ApJ...729...54C}
}

@ARTICLE{Cahoy2010,
    author = {{Cahoy}, Kerri L. and {Marley}, Mark S. and {Fortney}, Jonathan J.},
    title = "{Exoplanet Albedo Spectra and Colors as a Function of Planet Phase, Separation, and Metallicity}",
    journal = {The Astrophysical Journal},
    year = {2010},
    month = nov,
    volume = {724},
    number = {1},
    pages = {189--214},
    doi = {10.1088/0004-637X/724/1/189},
    adsurl = {https://ui.adsabs.harvard.edu/abs/2010ApJ...724..189C}
}

@ARTICLE{Freedman2014,
    author = {{Freedman}, Richard S. and
              {Lustig-Yaeger}, Jacob and
              {Fortney}, Jonathan J. and
              {Lupu}, Roxana E. and
              {Marley}, Mark S. and
              {Lodders}, Katharina},
    title = "{Gaseous Mean Opacities for Giant Planet and Ultracool Dwarf Atmospheres over a Range of Metallicities and Temperatures}",
    journal = {The Astrophysical Journal Supplement Series},
    year = {2014},
    month = oct,
    volume = {214},
    number = {2},
    eid = {25},
    pages = {25},
    numpages = {14},
    doi = {10.1088/0067-0049/214/2/25},
    adsurl = {https://ui.adsabs.harvard.edu/abs/2014ApJS..214...25F}
}

@ARTICLE{Korenaga2023,
    author = {{Korenaga}, Jun},
    title = "{Rapid Solidification of Earth's Magma Ocean Limits Early Lunar Recession}",
    journal = {Icarus},
    year = {2023},
    month = aug,
    volume = {400},
    eid = {115564},
    pages = {115564},
    doi = {10.1016/j.icarus.2023.115564},
    adsurl = {https://ui.adsabs.harvard.edu/abs/2023Icar..40015564K}
}

@ARTICLE{ModirroustaGalian2025,
    author = {{Modirrousta-Galian}, Darius and
              {Korenaga}, Jun},
    title = "{On the Efficacy of Ocean Formation With a Primordial Hydrogen Atmosphere}",
    journal = {Journal of Geophysical Research: Planets},
    year = {2025},
    month = aug,
    volume = {130},
    number = {8},
    eid = {e2024JE008786},
    pages = {e2024JE008786},
    doi = {10.1029/2024JE008786},
    adsurl = {https://ui.adsabs.harvard.edu/abs/2025JGRE..13008786M}
}

@article{DahlStevenson2010,
  author  = {Dahl, Tais W. and Stevenson, David J.},
  title   = {Turbulent Mixing of Metal and Silicate during Planet Accretion---and Interpretation of the {Hf--W} Chronometer},
  journal = {Earth and Planetary Science Letters},
  year    = {2010},
  volume  = {295},
  number  = {1-2},
  pages   = {177--186},
  doi     = {10.1016/j.epsl.2010.03.038}
}

@article{Zhou2024,
  author  = {Zhou, You and Driscoll, Peter E. and Zhang, Mingming and Reinhardt, Christian and Meier, Thomas},
  title   = {A Scaling Relation for Core Heating by Giant Impacts and Implications for Dynamo Onset},
  journal = {Journal of Geophysical Research: Planets},
  year    = {2024},
  volume  = {129},
  number  = {5},
  pages   = {e2023JE008163},
  doi     = {10.1029/2023JE008163}
}

@article{ModirroustaGalian2026,
  author  = {Modirrousta-Galian, Darius and Korenaga, Jun},
  title   = {Retention of Primordial Atmospheres through Giant Impacts},
  journal = {Nature Astronomy},
  year    = {2026},
  doi     = {10.1038/s41550-026-03006-1}
}

@article{Sudarsky2000,
  author  = {Sudarsky, David and Burrows, Adam and Pinto, Philip},
  title   = {Albedo and Reflection Spectra of Extrasolar Giant Planets},
  journal = {The Astrophysical Journal},
  year    = {2000},
  volume  = {538},
  number  = {2},
  pages   = {885--903},
  doi     = {10.1086/309160}
}

@article{Zeng2016,
  author  = {Zeng, Li and Sasselov, Dimitar D. and Jacobsen, Stein B.},
  title   = {Mass--Radius Relation for Rocky Planets Based on {PREM}},
  journal = {The Astrophysical Journal},
  year    = {2016},
  volume  = {819},
  number  = {2},
  pages   = {127},
  doi     = {10.3847/0004-637X/819/2/127}
}

@article{Lichtenberg2021,
  author  = {Lichtenberg, Tim and Bower, Dan J. and Hammond, Mark
             and Boukrouche, Ryan and Sanan, Patrick
             and Tsai, Shang-Min and Pierrehumbert, Raymond T.},
  title   = {Vertically Resolved Magma Ocean--Protoatmosphere Evolution:
             {H2}, {H2O}, {CO2}, {CH4}, {CO}, {O2}, and {N2}
             as Primary Absorbers},
  journal = {Journal of Geophysical Research: Planets},
  year    = {2021},
  volume  = {126},
  number  = {2},
  pages   = {e2020JE006711},
  doi     = {10.1029/2020JE006711}
}

@article{MiyazakiKorenaga2022,
  author  = {Miyazaki, Yoshinori and Korenaga, Jun},
  title   = {Inefficient Water Degassing Inhibits Ocean Formation
             on Rocky Planets: An Insight from Self-Consistent
             Mantle Degassing Models},
  journal = {Astrobiology},
  year    = {2022},
  volume  = {22},
  number  = {6},
  pages   = {713--734},
  doi     = {10.1089/ast.2021.0126}
}

@article{ModirroustaGalian2023,
  author  = {Modirrousta-Galian, Darius and Korenaga, Jun},
  title   = {The Three Regimes of Atmospheric Evaporation
             for Super-Earths and Sub-Neptunes},
  journal = {The Astrophysical Journal},
  year    = {2023},
  volume  = {943},
  number  = {1},
  pages   = {11},
  doi     = {10.3847/1538-4357/ac9d34}
}

@article{Hayashi1979,
  author  = {Hayashi, Chushiro and Nakazawa, Kiyoshi and Mizuno, Hiroshi},
  title   = {Earth's Melting Due to the Blanketing Effect of the Primordial Dense Atmosphere},
  journal = {Earth and Planetary Science Letters},
  year    = {1979},
  volume  = {43},
  number  = {1},
  pages   = {22--28},
  doi     = {10.1016/0012-821X(79)90152-3}
}

@article{MatsuiAbe1986,
  author  = {Matsui, Takafumi and Abe, Yutaka},
  title   = {Evolution of an Impact-Induced Atmosphere and Magma Ocean on the Accreting Earth},
  journal = {Nature},
  year    = {1986},
  volume  = {319},
  pages   = {303--305},
  doi     = {10.1038/319303a0}
}

@article{Calder2026,
  author  = {Calder, Robb and Shorttle, Oliver and Nicholls, Harrison
             and Lichtenberg, Tim and Guimond, Claire Marie},
  title   = {Most Rocky Sub-{Neptunes} Are Molten: Mapping the Solidification
             Shoreline for Gas Dwarf Exoplanets},
  journal = {Monthly Notices of the Royal Astronomical Society},
  year    = {2026},
  volume  = {549},
  number  = {3},
  pages   = {stag1007},
  doi     = {10.1093/mnras/stag1007}
}

@article{IkomaHori2012,
  author  = {Ikoma, Masahiro and Hori, Yasunori},
  title   = {In Situ Accretion of Hydrogen-rich Atmospheres
             on Short-period Super-Earths:
             Implications for the {Kepler-11} Planets},
  journal = {The Astrophysical Journal},
  year    = {2012},
  volume  = {753},
  number  = {1},
  pages   = {66},
  doi     = {10.1088/0004-637X/753/1/66}
}

@article{ModirroustaGalian2026MassLoss,
  author  = {Modirrousta-Galian, Darius and Korenaga, Jun},
  title   = {Revising Core Powered Mass Loss: A Critical Assessment
             of the ``Energy Limited'' Argument},
  journal = {Icarus},
  year    = {2026},
  volume  = {444},
  pages   = {116827},
  doi     = {10.1016/j.icarus.2025.116827}
}

@incollection{DAngelo2010,
  author    = {D'Angelo, Gennaro and Durisen, Richard H. and
               Lissauer, Jack J.},
  title     = {Giant Planet Formation},
  booktitle = {Exoplanets},
  editor    = {Seager, Sara},
  publisher = {University of Arizona Press},
  address   = {Tucson, AZ},
  year      = {2010},
  pages     = {319--346},
  eprint    = {1006.5486},
  archivePrefix = {arXiv},
  primaryClass  = {astro-ph.EP}
}

@ARTICLE{Paxton2011,
    author = {Paxton, B. and Bildsten, L. and Dotter, A.
              and Herwig, F. and Lesaffre, P. and Timmes, F.},
    title = {{Modules for Experiments in Stellar Astrophysics (MESA)}},
    journal = {\apjs},
    year = {2011},
    volume = {192},
    pages = {3},
    doi = {10.1088/0067-0049/192/1/3},
    archivePrefix = {arXiv},
    eprint = {1009.1622},
    primaryClass = {astro-ph.SR},
    adsurl = {https://ui.adsabs.harvard.edu/abs/2011ApJS..192....3P}
}

@ARTICLE{Paxton2013,
    author = {Paxton, B. and Cantiello, M. and Arras, P.
              and Bildsten, L. and Brown, E. F. and Dotter, A.
              and Mankovich, C. and Montgomery, M. H.
              and Stello, D. and Timmes, F. X. and Townsend, R.},
    title = {{Modules for Experiments in Stellar Astrophysics
              (MESA): Planets, Oscillations, Rotation, and Massive Stars}},
    journal = {\apjs},
    year = {2013},
    volume = {208},
    pages = {4},
    doi = {10.1088/0067-0049/208/1/4},
    archivePrefix = {arXiv},
    eprint = {1301.0319},
    primaryClass = {astro-ph.SR},
    adsurl = {https://ui.adsabs.harvard.edu/abs/2013ApJS..208....4P}
}

@ARTICLE{Paxton2015,
    author = {Paxton, B. and Marchant, P. and Schwab, J.
              and Bauer, E. B. and Bildsten, L. and Cantiello, M.
              and Dessart, L. and Farmer, R. and Hu, H.
              and Langer, N. and Townsend, R. H. D.
              and Townsley, D. M. and Timmes, F. X.},
    title = {{Modules for Experiments in Stellar Astrophysics
              (MESA): Binaries, Pulsations, and Explosions}},
    journal = {\apjs},
    year = {2015},
    volume = {220},
    pages = {15},
    doi = {10.1088/0067-0049/220/1/15},
    archivePrefix = {arXiv},
    eprint = {1506.03146},
    primaryClass = {astro-ph.SR},
    adsurl = {https://ui.adsabs.harvard.edu/abs/2015ApJS..220...15P}
}

@ARTICLE{Paxton2018,
    author = {Paxton, B. and Schwab, J. and Bauer, E. B.
              and Bildsten, L. and Blinnikov, S. and Duffell, P.
              and Farmer, R. and Goldberg, J. A.
              and Marchant, P. and Sorokina, E. and Thoul, A.
              and Townsend, R. H. D. and Timmes, F. X.},
    title = {{Modules for Experiments in Stellar Astrophysics
              (MESA): Convective Boundaries, Element Diffusion,
              and Massive Star Explosions}},
    journal = {\apjs},
    year = {2018},
    volume = {234},
    pages = {34},
    doi = {10.3847/1538-4365/aaa5a8},
    archivePrefix = {arXiv},
    eprint = {1710.08424},
    primaryClass = {astro-ph.SR},
    adsurl = {https://ui.adsabs.harvard.edu/abs/2018ApJS..234...34P}
}

@ARTICLE{Paxton2019,
    author = {Paxton, Bill and Smolec, R. and Schwab, Josiah
              and Gautschy, A. and Bildsten, Lars
              and Cantiello, Matteo and Dotter, Aaron
              and Farmer, R. and Goldberg, Jared A.
              and Jermyn, Adam S. and Kanbur, S. M.
              and Marchant, Pablo and Thoul, Anne
              and Townsend, Richard H. D. and Wolf, William M.
              and Zhang, Michael and Timmes, F. X.},
    title = {{Modules for Experiments in Stellar Astrophysics
              (MESA): Pulsating Variable Stars, Rotation,
              Convective Boundaries, and Energy Conservation}},
    journal = {\apjs},
    year = {2019},
    volume = {243},
    pages = {10},
    doi = {10.3847/1538-4365/ab2241},
    archivePrefix = {arXiv},
    eprint = {1903.01426},
    primaryClass = {astro-ph.SR},
    adsurl = {https://ui.adsabs.harvard.edu/abs/2019ApJS..243...10P}
}

@ARTICLE{Jermyn2023,
    author = {Jermyn, Adam S. and Bauer, Evan B.
              and Schwab, Josiah and Farmer, R.
              and Ball, Warrick H. and Bellinger, Earl P.
              and Dotter, Aaron and Joyce, Meridith
              and Marchant, Pablo and Mombarg, Joey S. G.
              and Wolf, William M. and Sunny Wong, Tin Long
              and Cinquegrana, Giulia C. and Farrell, Eoin
              and Smolec, R. and Thoul, Anne
              and Cantiello, Matteo and Herwig, Falk
              and Toloza, Odette and Bildsten, Lars
              and Townsend, Richard H. D. and Timmes, F. X.},
    title = {{Modules for Experiments in Stellar Astrophysics
              (MESA): Time-dependent Convection, Energy Conservation,
              Automatic Differentiation, and Infrastructure}},
    journal = {\apjs},
    year = {2023},
    volume = {265},
    pages = {15},
    doi = {10.3847/1538-4365/acae8d},
    archivePrefix = {arXiv},
    eprint = {2208.03651},
    primaryClass = {astro-ph.SR},
    adsurl = {https://ui.adsabs.harvard.edu/abs/2023ApJS..265...15J}
}

@misc{NASAExoplanetArchive2026,
    doi = {10.26133/NEA12},
    url = {https://doi.org/10.26133/NEA12},
    author = {{NASA Exoplanet Archive}},
    title = {Planetary Systems},
    publisher = {NExScI-Caltech/IPAC},
    version = {Version: 2026-10-27},
    year = {2026}
}
\bibliographystyle{aasjournalv7}

\end{document}